\documentclass[aps,pra,twocolumn,showpacs,showkeys,superscriptaddress,longbibliography,nofootinbib,floatfix]{revtex4-1}
\usepackage[utf8]{inputenc}
\usepackage[english]{babel}

\usepackage{amsmath,amssymb}
\usepackage{dsfont}
\usepackage{relsize}
\usepackage{xspace}
\usepackage{amsfonts}
\usepackage{amsmath}

\usepackage{graphicx}

\usepackage[colorlinks=true,allcolors=blue]{hyperref}
\usepackage{verbatim}

\usepackage{bbm}

\usepackage{color}
\definecolor{blue}{rgb}{0.2,0.2,0.8}

\DeclareMathOperator{\re}{Re}
\DeclareMathOperator{\im}{Im}
\DeclareMathOperator{\tr}{tr}

\DeclareMathOperator{\grad}{grad}

\renewcommand{\vec}{\operatorname{vec}}
\newcommand{\unity}{\ensuremath{{\rm 1 \negthickspace l}{}}}

\newcommand{\wrt}[1]{\:\mathrm{d}#1\:} 
\newcommand{\hc}{\text{h.c.}}  

\newcommand{\be}{\begin{equation}}
\newcommand{\ee}{\end{equation}}

\newcommand{\bpm}{\begin{pmatrix}}
\newcommand{\epm}{\end{pmatrix}}

\newcommand{\I}{\mathbbm{1}} 

\newcommand{\Leps}{\ensuremath{\mathlarger \varepsilon}}  
\newcommand{\Partial}[2]{\ensuremath{\frac{\partial{#1}}{\partial{#2}}}\,\,\xspace}

\newcommand{\ket}[1]{\ensuremath{| #1 \rangle}}
\newcommand{\bra}[1]{\ensuremath{\langle #1 |}}
\newcommand{\braket}[2]{\ensuremath{\langle #1|#2 \rangle}}
\newcommand{\ketbra}[2]{\ket{#1}\bra{#2}}

\newcommand{\expect}[1]{\ensuremath{\left\langle #1 \right\rangle}}

\newcommand{\sref}[1]{Sec.~\ref{#1}}
\newcommand{\aref}[1]{Appendix~\ref{#1}}

\newcommand{\lecocqII}{\textbf{Set~1}}
\newcommand{\lecocqIII}{\textbf{Set~2}}

\newcommand{\wc}{\omega_c}  
\newcommand{\wa}{\omega_a}  
\newcommand{\wl}{\omega_L}  
\renewcommand{\wr}{\omega_R}  
\newcommand{\gac}{g_{ac}}   
\newcommand{\gco}{g_{0}}    
\newcommand{\ac}{R(t)}  
\newcommand{\ha}{\hat{a}}
\newcommand{\hb}{\hat{b}}

\newcommand{\hs}{\hat{\sigma}_+}
\newcommand{\Hco}{H_\text{om}}    
\newcommand{\Ha}{H_\text{atom}}   
\newcommand{\nn}{\bar{n}}  

\begin{document}
\title{Optimal Control of Hybrid Optomechanical Systems for Generating Non-Classical States of Mechanical Motion}
\author{Ville Bergholm} \email{ville.bergholm@iki.fi} \affiliation{Dept. Chemistry, Technical University of Munich (TUM),  85747 Garching, Germany 
and\\
   Munich Centre for Quantum Science and Technology (MCQST),  Schellingstr.~4, 80799~M{\"u}nchen, Germany}
\author{Witlef Wieczorek} \email{witlef.wieczorek@chalmers.se} \affiliation{Department of Microtechnology and Nanoscience (MC2), Chalmers University of Technology, 41296 G\"oteborg, Sweden}
\author{Thomas Schulte-Herbr{\"u}ggen} \email{tosh@tum.de} \affiliation{Dept. Chemistry, Technical University of Munich (TUM),  85747 Garching, Germany and\\
   Munich Centre for Quantum Science and Technology (MCQST),  Schellingstr.~4, 80799~M{\"u}nchen, Germany}
\author{Michael Keyl} \email{michaelkeyl137@gmail.com} \affiliation{Dahlem Centre for Complex Quantum Systems, Free University of Berlin, 14195 Berlin, Germany}

\date{\today}

\begin{abstract}
Cavity optomechanical systems are one of the leading experimental platforms for controlling mechanical motion in the quantum regime. We exemplify that the control over cavity optomechanical systems greatly increases by coupling the cavity also to a two-level system, thereby creating a hybrid optomechanical system. If the two-level system can be driven largely independently of the cavity, we show that the non-linearity thus introduced enables us to steer the extended system to non-classical target states of the mechanical oscillator with Wigner functions exhibiting significant negative regions. We illustrate how to use optimal control techniques beyond the linear regime to drive the hybrid system from the near ground state into a Fock target state of the mechanical oscillator. We base our numerical optimization on realistic experimental parameters for exemplifying how optimal control enables the preparation of decidedly non-classical target states, where naive control schemes fail. Our results thus pave the way for applying the toolbox of optimal control in hybrid optomechanical systems for generating non-classical mechanical states.
\end{abstract}

\maketitle

\section{Introduction}

In view of quantum technologies (see, e.g., \cite{DowMil03}), optimal control techniques provide an increasingly useful toolbox to take quantum hardware to the limits of reaching target states with high fidelity, precision, sensitivity and robustness---one prominent example being feedback stabilization of predefined photon-number states in a box~\cite{Haroche11,Haroche13}. Systematic strategies to unlock and exploit the hardware potential in experimental settings with the help of optimal control methods can be found in a quantum control roadmap~\cite{control-roadmap2015}.

A recent physical system that has been added to the family of quantum hardware are mechanical resonators \cite{schwab_putting_2005,poot_mechanical_2012,optomech2014}. Pioneering experiments realized cooling of mechanical motion to the quantum ground state by direct cryogenic \cite{oconnell_quantum_2010} or by laser-based cooling techniques \cite{chan_laser_2011,teufel_sideband_2011}. A current focus lies on generating non-classical mechanical states, which are required to fully leverage mechanical resonators for applications in quantum metrology \cite{geraci_short-range_2010,johnsson_macroscopic_2016}, as quantum transducers \cite{stannigel_optomechanical_2010,safavi-naeini_proposal_2011} or for fundamental tests of quantum mechanics \cite{marshall_towards_2003,romero-isart_quantum_2011,schmole_micromechanical_2016}. Along these lines, the control over excitations of single phonons has been demonstrated by coupling mechanical motion to artificial atoms \cite{oconnell_quantum_2010,chu_quantum_2017} or to cavity light fields \cite{riedinger_non-classical_2016,hong_hanbury_2017}.

Cavity optomechanical systems constitute a successful platform for quantum control of mechanical motion~\cite{optomech2014}.
Importantly, the optomechanical interaction is intrinsically non-linear and, thus, would lend itself for directly generating non-classical states of mechanical motion \cite{bose_preparation_1997,mancini_ponderomotive_1997,marshall_towards_2003}. However, in real-world physical realizations \cite{chan_laser_2011,teufel_circuit_2011}, the single-photon strong coupling regime \cite{rabl_photon_2011,nunnenkamp_single-photon_2011} required for exploiting this non-linearity has not been achieved to date, with notable exceptions in cold atom optomechanics setups \cite{murch_observation_2008,brennecke_cavity_2008}.
Therefore, it is common practice to boost the optomechanical interaction by a coherent drive at the cost of losing its intrinsic non-linear character.
Non-classical mechanical states can nevertheless be generated when coupling an optomechanical system to other non-linear systems \cite{hammerer_strong_2009,oconnell_quantum_2010,pflanzer_optomechanics_2013,ramos_nonlinear_2013}, by introducing the non-linearity in the measurement process \cite{jacobs_engineering_2009,vanner_quantum_2013,riedinger_non-classical_2016,clarke_growing_2019}, by injecting non-classical states of light \cite{akram_single-photon_2010,rabl_photon_2011,nunnenkamp_single-photon_2011,riedinger_non-classical_2016,hong_hanbury_2017}, or by coupling to the squared mechanical position quadrature \cite{thompson_strong_2008,sankey_strong_2010,vanner_selective_2011,romero-isart_large_2011}.

\begin{figure}[ht!]
  \includegraphics[width=1\columnwidth]{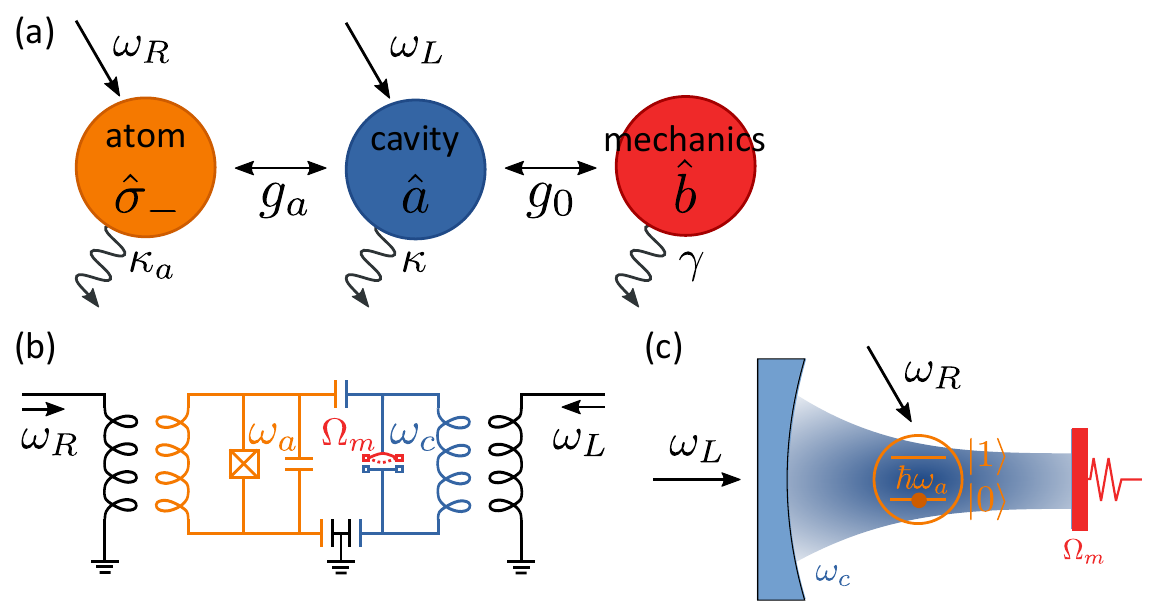}
  \caption{\label{fig:schematic} Overview of the hybrid optomechanical system analyzed. (a) We consider the interaction of a mechanical resonator (represented by its annihilation operator $\hat{b}$) with a cavity field (represented by its annihilation operator $\hat{a}$), which is at the same time coupled to an atom (represented by its lowering operator $\hat{\sigma}_-$). Possible physical implementations of this conceptual model can be realized in (b) an electromechanical circuit QED architecture (see Ref. \cite{lecocq_resolving_2015,schmidt_ultrawide-range_2018}) or (c) in an optomechanical cavity which simultaneously traps an atom.}
\end{figure}

In the present work we focus on a hybrid optomechanical system for non-classical state generation. More precisely, we choose a system consisting of a mechanical resonator that is parametrically and weakly coupled to a cavity, which in turn is strongly coupled to a two-level system (see Fig.~\ref{fig:schematic}). Such a system has been analyzed before in the context of strong atom-mechanics coupling \cite{hammerer_strong_2009}, dissipative state engineering \cite{pflanzer_optomechanics_2013}, tripartite polaron dynamics \cite{restrepo_single-polariton_2014}, and optical bistability \cite{jiang_optical_2016}. It finds direct relevance in present experimental implementations \cite{lecocq_resolving_2015,schmidt_ultrawide-range_2018} and potential future implementations in nano-optic \cite{tiecke_nanophotonic_2014,chan_laser_2011} or ion-trap scenarios \cite{neumeier_reaching_2018}. 

In our study, we first establish controllability of the optomechanical hybrid system in the absence of dissipation processes. We then apply optimal control algorithms and suggest concrete controls in an experimentally realistic setting for generating a non-classical state of mechanical motion. As an example, we choose to focus on generating a single-phonon Fock state and use numerical optimization to find pulse sequences for optimally generating such a state. In order to be close to realistic experimental
settings, we adapt parameters from the electromechanics implementation of Ref.~\cite{lecocq_resolving_2015} for illustrating the gain of optimal control over established control techniques.

In many instances, quantum optimal control \cite{Sam90+, peirce_optimal_1988, Krotov, Kha+05, dalessandro_introduction_2007} provides both framework and algorithms to go beyond conventional approaches. In cavity optomechanics, standard control techniques imply making use of interactions in the linearized regime using cw-driving \cite{vitali_optomechanical_2007,paternostro_creating_2007,wilson-rae_theory_2007,marquardt_quantum_2007,hofer_quantum_2011,hofer_entanglement-enhanced_2015,HoferHammerer_2017}, multi-tone driving \cite{kronwald_arbitrarily_2013,brunelli_unconditional_2018,houhou_unconditional_2018} or pulsed driving \cite{vanner_pulsed_2011,vanner_selective_2011,vanner_quantum_2013} for generating entangled states \cite{palomaki_entangling_2013,riedinger_remote_2018,ockeloen-korppi_stabilized_2018}, squeezed states \cite{wollman_quantum_2015,pirkkalainen_squeezing_2015,lecocq_quantum_2015} or for performing state tomography \cite{vanner_cooling-by-measurement_2013}. The leap that optimal control techniques offer is to accommodate the specifics of the system for finding experimental protocols that can be run in a much shorter time frame or that achieve a higher fidelity in state preparation. 
Optimal control may, thus, find control sequences that embrace limitations in the system's parameters, which otherwise may prevent generating desired target states. Indeed, optimal control schemes have already been analyzed for optomechanical systems for enhancing cooling performance \cite{wang_ultraefficient_2011,machnes_pulsed_2012,triana_ultrafast_2016}, for generating optomechanical entanglement \cite{stefanatos_maximising_2017}, or for squeezing~\cite{Koch_Squeezing_2018}. A recent account on treating optomechanical systems 
including feedback as {\em linear control systems} can be found in~\cite{HoferHammerer_2017}. 

In the present work, we move to the framework of {\em bilinear control systems}~\cite{Elliott09} and cutting-edge algorithms~\cite{machnes_2011} to show how (by adding a two-level atom) this setting allows for generating a non-classical mechanical Fock state in a parameter regime, where conventional steering methods fail. Our methodology can be extended for optimal generation of mechanical Schr\"odinger cat states \cite{mancini_ponderomotive_1997,brunelli_unconditional_2018} or cubic phase states \cite{houhou_unconditional_2018}, provided the truncation of the Hilbert space required for our computational optimization can be extended to higher Fock state numbers. 

Our paper is structured as follows: in \sref{sec:model} we present the Hamiltonian of the hybrid optomechanical system and derive a drift and control part, in \sref{sec:controllability} we discuss the controllability of our system from a general perspective, in \sref{sec:numerical-algorithms} we present the numerical algorithms our optimal control optimization is based on, in \sref{sec:simulation-results} we use optimal control based on realistic experimental parameters to generate a single phonon Fock state and finally, in \sref{sec:discussion}, we discuss the results and give an outlook on future work.
Appendix \ref{sec:deriv-optom-hamilt} summarizes a detailed derivation of the Hamiltonian
of the hybrid optomechanical system treated here and lists the entire parameter setting.

\section{Theoretical models: drift and control Hamiltonian}\label{sec:model}

The optomechanical system of interest is described in the lab frame by the Hamiltonian
\be
\begin{aligned}
\label{eq:om:co_}
\Hco/\hbar \;=\;& \wc \ha^\dagger \ha +\Omega_m \hb^\dagger \hb -\gco \ha^\dagger \ha (\hb+\hb^\dagger)\\
&+E(t) \cos(\wl t +\phi_L(t)) (\ha+\ha^\dagger),
\end{aligned}
\ee
where $\ha$ and $\hb$ are the annihilation operators of the cavity and the oscillator,
and the last term represents driving of the optical cavity by a laser.
The $\hat{Q} = \ha +\ha^\dagger$ quadrature of the cavity is defined as the direction of the driving.
The driving Rabi frequency is connected to the laser power~$P$ and the
cavity decay rate $\kappa$ by $E = \sqrt{\frac{2\kappa P}{\hbar \wl}}$.

The system may be made more amenable to control by adding a strongly non-linear element
in the form of a controllable two-level atom in the cavity with the Hamiltonian
\be
\begin{aligned}
\label{eq:om:atom_}
\Ha/\hbar
\;=\;&
\wa \hat{\sigma}_+ \hat{\sigma}_-
+\gac (\ha e^{i\phi_c}  +\hc)(\hs e^{i\phi_a} +\hc)\\
&+\ac \cos(\wr t +\phi_R(t))(\hs +\hc).
\end{aligned}
\ee
Above, the three terms represent the atom itself, the atom-cavity coupling,
and a classical control signal driving the atom,
respectively. $\hat{\sigma}_\pm$ denote the atomic raising and lowering
operators.

Dissipation processes to be taken into account are decay processes in
the cavity and damping of the mechanical oscillator. The former are described
by the Lindblad operator $\hat{V}_1 = \sqrt{\kappa} \ha$.
This assumes that the effective temperature of the cavity surroundings is zero,
which is a good approximation for most cavities.
To describe the damping in the mechanical oscillator we use the Lindblad
operators
$\hat{V}_2 = \sqrt{\gamma'}\hb$ and $\hat{V}_3 = \sqrt{\gamma' x}\hb^\dagger$, where $\gamma' = \gamma (\nn+1)$ is the effective decay rate,
$\gamma$ the base decay rate,
$\nn = x/(1-x)$ the expected number of oscillator phonons in the steady state,
and $x = e^{-\frac{\hbar\Omega_m}{kT}}$ the oscillator Boltzmann factor. Finally, we describe the atomic decay by
$\hat{V}_4 = \sqrt{\kappa_a} \hat{\sigma}_-$ with the atom decay rate
$\kappa_a$.  Combining all dissipation processes we end up with a standard Markovian master equation
\begin{equation} \label{eq:1}
  \frac{d}{dt} \rho = \frac{-i}{\hbar} [\Hco+\Ha,\rho] +
  \sum_{i=1}^4 \Big( \hat{V}_i\rho \hat{V}_i^{\dagger} - \frac{1}{2}
    \{\hat{V}_i^\dagger \hat{V}_i, \rho\}\Big),  
\end{equation}
where $\rho$ is the density operator of the overall system.

To get a set of equations better prone for numerical
simulation we simplify the described setup in  a number of steps
(cf. Appendix \ref{sec:deriv-optom-hamilt} for detailed calculations):
\begin{enumerate}
\item
  We transform the cavity into a frame co-rotating with the
  laser by applying the unitary $\exp(i t \wl
  \hat{a}^\dagger\hat{a})$. This is followed by a rotating wave
  approximation (RWA) to drop the counter-rotating terms.
\item
  Simultaneously, we transform the atom by $\exp(i t \wr
  \hat{\sigma}_+ \hat{\sigma}_-)$ and apply again the RWA.
\item
  Due to driving the cavity with a laser field, the cavity (without atom,
  and oscillator) ends up in a coherent steady state
  $|\alpha\rangle$. From a computational point of view it is useful to
  consider oscillations around $|\alpha\rangle$ (and not around the
  cavity vacuum), since this step allows more radical truncations of
  the physical Hilbert space.  Hence we apply a phase space shift to
  get new creation  and annihilation operators $a,a^\dagger$,
  $b,b^\dagger$ with 
  \[ \hat{a} = e^{i(\eta-\phi_L)} (a + s \I),\quad \hat{b} =
    e^{i\zeta}(b + r \I) \]
  with appropriately chosen parameters $\eta, s, \zeta,
  r$; cf. Appendix~\ref{sec:op_shift} for exact values.
  The $s$ shift, in particular, will act as a multiplicative factor to $\gco$
  in a new linear interaction term coupling the cavity and the oscillator.
\item
  The atomic operators $\hat{\sigma}_\pm$ are replaced by the phase-rotated
  versions $\sigma_\pm$ with
  \[ \sigma_+ = e^{i \phi} \hat{\sigma}_+, \quad \phi = \phi_a +\phi_c +\eta -\phi_L.\]
\item 
  If the average photon number is high, the non-linear optomechanical  interaction
  term can be linearized and replaced by a hopping interaction.
\item
  At the end we drop another counter-rotating $ab$ interaction term
  (i.e. another RWA), to make the system completely time independent
  (apart from the control terms).
\end{enumerate}
Finally, we end up with the following drift and control Hamiltonians:
\begin{align}
\label{eqn:H0-terms} 
H_\text{drift}/\hbar
=
(\wc'-&\wr) (a^\dagger a +b^\dagger b)
+({\wa}_0-\wr) \sigma_+ \sigma_-\\
\notag
&-\gco s \: (ab^\dagger +\hc) +\gac (a\sigma_+ +\hc),\\
\notag
H_\text{control}(t)/\hbar =&
u_{\text{detuning}}(t) \: 2\pi \sigma_+ \sigma_-
+u_{\text{atomX}}(t) \: \pi (\sigma_++\sigma_-)\\
&+u_{\text{atomY}}(t) \: \pi (-i)(\sigma_+-\sigma_-), \label{eqn:Hc-terms}
\end{align}
and with the modified Lindblad operators
\begin{equation}\label{eqn:L-terms}
  V_1 = \sqrt{\kappa}a,\; V_2 = \sqrt{\gamma'}b,\; V_3 =
  \sqrt{\gamma' x}b^\dagger,\; V_4 =  \sqrt{\kappa_a} \sigma_-.
\end{equation}

Now we can replace the Hamiltonian part of Eq.~\eqref{eq:1} by
$H_{\mathrm{tot}}(t):=H_{\mathrm{drift}}+H_{\mathrm{control}}(t)$ 
and the $\hat{V}_i$ by the $V_i$ of Eq.~\eqref{eqn:L-terms} 
to get a new master equation
\begin{align}\label{eqn:L-Master}
  \frac{d}{dt} \rho &= \frac{-i}{\hbar} [H_{\mathrm{tot}}(t), \rho] + \Gamma(\rho) \nonumber \\
	&\text{with}\quad \Gamma(\rho):=\sum_{i=1}^4 \big( V_i\rho V_i^{\dagger} - \tfrac{1}{2}
  \{V_i^\dagger V_i, \rho\}\big)\;.
\end{align}
Its time dependence is entirely wrapped up in the control functions
$u_j(t)$ with $j\in\{\text{{\small atomX, atomY, detuning}}\}$. In principle one could also control the frequency $\wl=\wl(t)$ and amplitude $E=E(t)$ of the laser drive
(see, e.g.,~\cite{machnes_pulsed_2012}) thus leading to a modulation of the cavity-laser detuning $\Delta=\Delta(t)$ 
and the cavity-oscillator coupling $g_0s=g_0s(t)$. 
For simplicity, in this work we keep the cavity laser drive constant in order to exploit the drive continuously for three tasks: 
(i)~cooling of the oscillator into its ground state, 
(ii)~using the swap interaction between cavity and oscillator and 
(iii)~separating the interaction time scales between the atom-cavity and the cavity-oscillator couplings. 
Yet, other parameter regimes or target states may require control of the external laser drive, which we leave for future work.

In our case, the optimal control task now amounts to finding control amplitudes $u_j(t)$ 
such that an appropriately chosen initial state $\rho_0$ evolves after
a time~$T$ into some best approximation to a given target state
$\rho_T$ of the mechanical oscillator (after tracing out cavity and atom). An obvious candidate for the initial state $\rho_0$ is
the steady state the system evolves into if we only consider laser driving of the
cavity. The steady state is influenced by the presence of the atom
and thus changes with the detuning of the atom. If the
detuning is large, $\rho_0$ is close to the ground state.
Numerical calculations can be found in Appendix~\ref{sec:steady_state}.

\section{Controllability}\label{sec:controllability}

Before analysing an experimentally realistic scenario,
let us sketch that asking for controllability of the hybrid optomechanical system proposed is
well-posed from a control theoretical point of view. To this end, we neglect
dissipation for the moment and solely look at the coherent evolution given by
$H_{\mathrm{drift}}+H_{\mathrm{control}}(t)$.

It is well known that the extent of control over harmonic oscillator modes or light modes greatly
increases by coupling the modes to a controllable two-level atom,
whereby the system actually becomes fully controllable \cite{LE96,brockett2003controllability,rangan2004control}. 
For the Jaynes-Cummings model of one or several atoms 
coupled to an oscillator mode,  some of us  showed in \cite{KZSH} that breaking the 
symmetry $\sigma_z\otimes N$ by controls on the atom leads to
approximate full controllability (in the strong operator topology).
Here $\sigma_z$ acts on the atom and $N$ is the number operator of the oscillator.

Systematically extending similar lines, approximate controllability of the
Jaynes-Cum\-mings-Hubbard model (now comprising an entire network of cavities each 
containing one mode and one two-level atom, where the interaction between two cavities 
is given by a hopping term)
is analyzed in detail in~\cite{HeKe18}. It is shown that any pure state of the overall system can be
prepared with arbitrarily small error starting from an arbitrary pure
initial state by controlling the atoms individually and the
cavity-cavity interactions globally. 
Yet when looking at the mechanical
oscillator as another cavity, this result does not directly cover
our case, because the oscillator in turn is not coupled to another
controllable two-level system. Moreover, the reasoning in \cite{HeKe18} indicates
that the given scenario is a minimal requirement for full controllability. 
Hence, in our case this means one could not prepare any pure state of the
\emph{overall system} either.

Fortunately, the overall state of the system is not needed, since in the extended system
suggested here, we are interested in the {\em partial state} of the oscillator only, where the situation 
is much easier. Consider the atom-cavity subsystem first. 
This part is well studied and known to be fully controllable
\cite{brockett2003controllability,rangan2004control,yuan2007controllability,bloch2010finite,KZSH}
in the sense that one can reach with arbitrary small error any pure state of atom and cavity 
from any initial state by using $u_{\text{atomX}}(t)$ and $u_{\text{atomY}}(t)$ 
only---with $u_{\text{detuning}}(t)$ not needed except for speed-up. 

In contrast, cavity and oscillator alone just follow quasi-free dynamics, which
is much more limited. Even when allowing for control of the coupling strength, 
one is far from full controllability, since the time-evolution 
operator would be confined to the Schr\"odinger representation of the metaplectic group
\cite{folland2016harmonic}. However, with the Hamiltonian components given in 
Eqs.~\eqref{eqn:H0-terms} and~\eqref{eqn:Hc-terms},
it is easy to see that---in entire analogy to the classical phase space---the 
states of cavity and oscillator become (approximately)
flipped after a certain time. Hence a possible strategy to control the
overall system is: prepare a state of the cavity first and wait
until it flips over to the oscillator. This indicates that one
can actually prepare \emph{any} pure state of the oscillator from an
arbitrary pure initial state.

In this idealized controllability assessment of our setup, note that
atom-cavity coupling $\gac$ is 10 times stronger than the boosted
cavity-oscillator coupling~$\gco s$ thus leading to a separation of
time scales. In simple cases this allows to treat atom-cavity and cavity-oscillator system
independently as described above. 
In other words, a
simple cavity state can be prepared before the swapping to the
oscillator has effectively started. After this preparation
phase, the effects of the still present atom-cavity interaction can be
continuously compensated by further control pulses on the atom
(such that the swapping can go on undisturbed). This approximation
only breaks down if the preparation of the cavity state takes so long
that it gets compromised by the cavity-oscillator swap. In that case it
may be necessary to resort to controlling the parameter $s$ in order to
manipulate the cavity-oscillator coupling~$\gco s$. Another limitation
to the controllability assessment just outlined is dissipation. So both the numerical
analysis and the experiment have to address mixed states with the
dissipation time limiting the overall control time.

The state-of-the-art of using optimal control with linear feedback for
optomechanical systems has been summarized in the recent comprehensive
review by Hofer and Hammerer~\cite{HoferHammerer_2017}. Note that the
systems thus far addressed do not use the interaction
with an atom, but rather a feedback loop from homodyne detection on a beam
coupled out of the cavity. The information gathered is then used to drive
the cavity with a {\em linear} feedback Hamiltonian $H_{\mathrm{fb}}$ of the form
\[
H_{\mathrm{fb}}/\hbar = - (\epsilon(t)^* a + \epsilon(t) a^\dagger),
\]
where $a, a^{\dagger}$ are annihilation and creation operators of the
cavity and $\epsilon(t) \in \mathbb{C}$ is the amplitude of the feedback
signal. From the point of view of Lie-algebraic systems and control theory,
such systems  come with limitations: since all
terms in the overall Hamiltonian are at most quadratic in creation and
annihilation operators, Hamiltonians of that form constitute a \emph{finite
dimensional} Lie algebra. Therefore, the manifold of time evolution
operators thus generated is also finite dimensional (no matter whether the terms
are time dependent or not) and can thus be described by finitely many parameters. 
A given initial state with Wigner function $W_0$ evolves following a classical
phase flow, i.e.~the solution of the initial value problem for the
classical system. The latter, however, is but a multi-dimensional
harmonic oscillator driven by a force which is constant in space.
Hence, if $W_0$ has no negative parts, this cannot change under such a form of 
time evolution. For instance, if $W_0$ is Gaussian, it stays Gaussian all the time.
 
Adding an atom interacting with the cavity as used in our context is meant to overcome 
exactly these limitations.
One may look at it as adding a third oscillator which only
interacts via its two lowest levels with the rest of the system. The
corresponding interaction term cannot be written as a quadratic
polynomial in creation and annihilation operators of the now
three-dimensional oscillator system. Therefore it breaks the
covariance of the canonical commutation relations given in
terms of the metaplectic representation, and the reasoning from the
previous paragraph does not apply. Therefore, adding a
two-level atom allows for preparing any state of the harmonic
oscillator subsystem from any initial state.---The remaining question of 
how severe the restrictions imposed by a
{\em realistic dissipative system} are,  and up to which degree the
theoretical possibilities can actually be exploited by pulse sequences
shall be explored in the sequel  by some examples using numerical
optimal control.

\section{Numerical algorithms}
\label{sec:numerical-algorithms}
In view of going beyond Gaussian states, the extended hybrid
optomechanical setting lends itself to be treated as a {\em bilinear control system}~\cite{Elliott09}
with {\em states} $X(t)$ following
\begin{equation}
\dot X(t) = \big(A+\sum_j u_j(t) B_j\big)\, X(t)\quad\text{with}\quad X(0)=X_0\;.
\end{equation}
Its form is determined by a non-switchable {\em drift term} $A$, 
while the control is brought about by (typically piecewise constant) {\em control amplitudes} $u_j(t)\in\mathbb R$ 
governing the time dependence of the otherwise constant {\em control operators} $B_j$. 
The connection to the Lindblad master Eq.~\eqref{eqn:L-Master} above is given by the identifications\footnote{
  Here, $\hat{H}_{\sf drift}$, $\hat{H}_{\sf control}$ and $\hat\Gamma$ are linear \/`superoperators\/'
  acting on the state $\rho(t)$. For numerics, a convenient concrete representation takes $\rho$ as the column vector $\vec(\rho)$ stacking
  all columns of the matrix~$\rho$. With the conventions of Ref.~\cite[Chp.~4]{HJ2}, Hamiltonian commutator superoperator components $\hat H_j$
  are obtained as
  $\hat H_j:= (\unity\otimes H_j - H_j^\top\otimes\unity)$, and the Lindbladian dissipator as
  $\hat\Gamma:= \sum_k\bar V_k\otimes V_k -\tfrac{1}{2}\big(\unity\otimes (V_k^\dagger V_k) + (V_k^\top\bar V_k)\otimes \unity\big)$.
  The Lindblad master eqn.~\eqref{eqn:L-Master} can
  then readily be read and treated as vector differential equation of the type $\vec(\dot\rho)=(-\tfrac{i}{\hbar} \hat H + \hat \Gamma) \vec(\rho)$.}
\begin{align}
\rho(t) &=: X(t) \\
\hat{\Gamma} -i \hat{H}_{\sf drift}/\hbar &=: A \qquad\text{(Eqs.~\eqref{eqn:H0-terms} and~\eqref{eqn:L-terms})} \\
-i\hat{H}_{\sf control}(t)/\hbar &=: \sum_j u_j(t) B_j \quad\text{(Eq.~\eqref{eqn:Hc-terms})}\;.
\end{align}

Given this equation of motion, the optimal control task then amounts to minimizing the Euclidean distance
between the (possibly mixed) target state $\rho_T$ on the one hand and the final state $\rho(T) $ of the system on the other hand.
Typically $\rho(T)$ results after $n$ steps of time propagation in slots of piecewise constant quantum maps $\hat F_k$ 
(with $\tau:=t_k-t_{k-1}$ for $k=2,\dots,n$ as uniform width of time intervals) propagating the state $\rho(0)$ according to Eqn.~\eqref{eqn:L-Master}
\begin{align}
\rho(T) &\phantom{:}= \hat F_n\circ \hat F_{n-1}\circ\dots\circ \hat F_k \circ\dots \circ \hat F_1 \; \rho(0),\;\text{where}\\
\hat F_k &:= e^{\tau \hat{L}_k}\quad\text{with}\quad \hat{L}_k:= \hat{\Gamma} - i\sum_j u_j(t_k) \hat{H}_j \; .
\end{align}
Likewise, the distance between the truncations to the sublevels of interest $\tr_E(\rho_T)$ and $\tr_E(\rho(T))$ may be taken,
or alternatively, a Lagrange-type penalty term may be added to the cost functionals discussed in the outlook. 

Explicitly allowing for changing purity and mixed target states requires some generalization of the standard task (with constant purity) discussed in 
Ref.~\cite{machnes_2011}. To this end, we extract from the (squared) Euclidean distance 
(in terms of the Frobenius norm $||A||_F:= \sqrt{\tr\{A^\dagger A\} }$)
\begin{equation}
D:=|| \rho_T - \rho(T)||^2_F = ||\rho_T ||^2_F + ||\rho(T) ||^2_F - 2 \re \tr \{\rho_T^\dagger \rho(T) \}
\end{equation}
those terms depending on time (and therefore on the controls) and rescale to arrive at the cost functional
\begin{equation}
\Leps := \tfrac{1}{2}\,||\rho(T)||^2_F  -  \re \tr \{\rho_T^\dagger \rho(T) \}\;.
\end{equation}
Taking the derivative with respect to the control amplitude $u_j$ in the $k^{\rm th}$ time slot then gives
\begin{widetext}
\begin{align}
\Partial{\Leps}{u_j(t_k)} &= \; \re \tr \Big\{\rho(T)^\dagger \Partial{}{u_j(t_k)} \rho(T)\Big\} - \re \tr \Big\{\rho_T^\dagger \Partial{}{u_j(t_k)} \rho(T) \Big\}\\[1mm]
	&= \; \re \, \tr \Big\{ \big(\rho(T) - \rho_T\big)^\dagger\;  \Partial{}{u_j(t_k)} \rho(T)\Big\}\\[1mm] 
	&= \; \re \, \tr \Big\{ \big(\rho(T) - \rho_T\big)^\dagger\; \hat F_n\circ \hat F_{n-1}\circ\dots\circ \hat F_{k+1} \circ \bigg(\Partial{\hat F_k}{u_j(t_k)}\bigg)\circ \hat F_{k-1}\circ\dots\circ \hat F_1\; \rho(0)\Big\}  \;,\quad\phantom{.}
\end{align}
\end{widetext}
where the difference $(\rho(T) - \rho_T)^\dagger$ instead of just $(-\rho_T)^\dagger$ now takes care of the purity change.
In the unital case, the derivative of the propagating quantum map $\hat F_k$
would make use of $\hat F_k$ being normal 
(so in slight abuse of language it has orthogonal eigenvectors $\ket{\lambda^{(k)}_i}$ associated to the real eigenvalues $\lambda^{(k)}_i$)
to take the form described in \cite{Aizu63, Wilcox67} and used in~\cite{machnes_2011}:
\begin{equation}
\hspace{-5mm}
\Partial{\hat F_k}{u_j(t_k)} = \begin{cases} 
	-\, \braket{\lambda^{(k)}_a} { i \hat{H}_j \lambda^{(k)}_b}\; \tau\, e^{\tau \lambda^{(k)}_a}    &\text{for} \; \lambda^{(k)}_a = \lambda^{(k)}_b \\[2mm]
	-\, \braket{\lambda^{(k)}_a} { i \hat{H}_j \lambda^{(k)}_b}\; \tfrac{ e^{\tau\lambda^{(k)}_a} - e^{\tau\lambda^{(k)}_b} }{\lambda^{(k)}_a - \lambda^{(k)}_b}   &\text{for} \; \lambda^{(k)}_a \neq \lambda^{(k)}_b
	\end{cases}.
\end{equation}
In the general (non-normal) case mostly encountered here we have to resort to finite differences according to
\begin{equation}
\Partial{\hat F_k}{u_j(t_k)} \simeq \frac{e^{\tau(\hat{L}(u_j(t_k)) -i \delta\hat{H}_j)} - e^{\tau \hat{L}(u_j(t_k))}}{\delta}\;,
\end{equation}
where $\delta$ has to be sufficiently small in the sense $\delta~\ll~1/||\tau\, \hat F_k||$.
Given $\Partial{\hat F_k}{u_j(t_k)}$,
steepest-descent of the cost functional with the controls would follow a recursion in $r$ reading 
\begin{equation}\label{eqn:recursion}
u_j^{(r+1)}(t_k) = u_j^{(r)}(t_k) + \alpha_r \frac{\partial \Leps^{(r)}}{\partial u_j(t_k)}
\end{equation}
with $\alpha_r$ as step size, while the standard Newton update would take the form
\begin{equation}
\ket{u^{(r+1)}(t_k)} = \ket{u^{(r)}(t_k)} + \alpha_r \mathcal{H}_r^{-1} \ket{\grad \Leps^{(r)}(t_k)}
\end{equation}
with $\mathcal{H}_r^{-1}$ denoting the
inverse Hessian in the $r^{\rm th}$ iteration. For convenience the array of piecewise constant control amplitudes 
$\{u_j^{(r)}(t_k)\,|\, j=1,2,\dots, m\}$ is concatenated to the control vector \ket{u^{(r)}(t_k)} for each time slot
$\{t_k\,|\, k=1,2,\dots, n\}$, while \ket{\grad \Leps^{(r)}} is the corresponding gradient vector.
In this work we use the {\sc bfgs} quasi-Newton algorithm~\cite{NocWri06} to approximate the inverse Hessian as explained in~\cite{machnes_2011}.

\section{Results by optimal control}
\label{sec:simulation-results}

The numerical optimization results presented in this section are for two variants of the circuit cavity electromechanical system described in~\cite{lecocq_resolving_2015}.
It consists of a mechanical oscillator coupled to a microwave cavity.
The cavity mode is further coupled to a superconducting qubit (``atom'').
In the original implementation the atom-cavity coupling is fairly strong,~$\gac/(2\pi) = 12.5$~MHz,
but the cavity-oscillator coupling is much weaker, $\gco/(2\pi)= 300$~Hz.
In all of our simulations we have artificially boosted the single-photon optomechanical coupling
strength $\gco$ by one order of magnitude,
which together with the boost~$s$ resulting from coherent driving of the cavity
brings the optomechanical system in the required strong-coupling regime\footnote{
  Note that one could in principle boost $s$ instead of $\gco$
  to reach the required strong coupling regime $s\cdot g_0>\kappa,\gamma$ between the mechanical oscillator and the cavity.
  However, boosting $s$ also increases the interaction of the atom with the cavity,
  which complicates the control scheme as discussed in \aref{app:controlsystem}.}.

The two parameter sets we use are
\begin{itemize}
\item
  \textbf{\lecocqII}:
  Coupling enhancement factor $s=100$, $\gco/(2\pi)$ is boosted by a factor of~$40$ to $12$~kHz, cavity decay rate $\kappa/(2\pi) = 1$~MHz and the device is operated at a temperature of $25\,$mK.
\item
  \textbf{\lecocqIII}:
  Coupling enhancement factor $s=120$, $\gco/(2\pi)$ is boosted by a factor of~$10$ to $3$~kHz,
  cavity decay rate $\kappa/(2\pi) = 0.2$~MHz and the device is operated at a  temperature of $10\,$mK.
  Compared to \textbf{\lecocqII}, we have assumed an optical cavity with a smaller linewidth, which allowed us to reduce
  the boost of the optomechanical coupling strength\footnote{
    If the $\gco$ coupling strength we propose turns out to be experimentally infeasible,
    our results indicate that the loss in controllability due to a lower $g_0$ can be compensated by further decreasing~$\kappa$.
    We are confident that one could reach a regime experimentally where $\gco$ is mildly increased beyond the value reported in~\cite{lecocq_resolving_2015},
    e.g.,~by reducing the gap between the plates of the capacitor, and at the same time the optical quality factor $\kappa$ of the microwave cavity is improved,
    e.g.,~by using a three-dimensional implementation~\cite{reagor_reaching_2013}.
  }.
  We have also assumed a dilution fridge operating at $10\,$mK.
\end{itemize}

A full list of parameters is found in the Appendix in Table~\ref{table:symbols},
along with related parameter ratios in Table~\ref{table:measures}.
In the following, we shortly discuss parameter ratios that are relevant for achieving quantum control as commonly known from optomechanical or cavity QED setups.
We require the optomechanical system to be sideband-resolved, i.e., $\Omega_m/\kappa>1$.
This facilitates efficient state swap of the cavity state to the mechanical oscillator by selecting
the beam-splitter (hopping) interaction from the optomechanical interaction Hamiltonian and at the same time
suppressing the undesired two-mode squeezing part of the Hamiltonian.
We need the optomechanical cooperativity to be larger than unity
($\frac{|\gco s|^2}{\kappa \, \gamma \, \nn}>1$), which allows the mechanical
oscillator to be laser-cooled close to the quantum ground state, the initial state of our system.
We need to be in the strong coupling regime, both for the cavity-oscillator part ($\frac{|\gco s|}{\max(\kappa, \gamma)}>1$)
as well as for the atom-cavity part ($\frac{\gac}{\max(\kappa, \kappa_a)}>1$).
The former facilitates coherent swapping of the state of the cavity to the oscillator,
and the latter from the atom to the cavity. All these conditions are fulfilled for all chosen parameter sets.

The system is controlled by driving the atomic transition harmonically (the {atomX} and {atomY} controls), and by
adjusting the atomic resonance frequency~$\wa$, which changes the detuning $\wa-\wr$ of the atom
from the driving signal (the {detuning} control).
The frequency $\wr$ of the driving signal is $2\pi \cdot 500$~MHz below the shifted cavity resonance frequency~$\wc'$,
which allows us to draw a clear separation between the atom being
resonant with the drive, or with the cavity, or neither.

At the beginning of the optimization each control field in the sequence is initialized to a random value.
The control sequence is first optimized for a short computational time (about $300$~s) without dissipation
to quickly obtain a reasonable starting sequence,
and then for a longer computational time (several hours) with the computationally heavier dissipation processes included.
Due to many local minima (typical of open-system optimization),
generically one has to repeat the optimization with random initial sequences dozens of times to obtain sufficiently good results.

We simulate the harmonic oscillator modes by truncating the infinite-dimensional Fock space into
a finite-dimensional one. To make sure our control sequences remain valid in the untruncated case,
we apply a penalty functional on the population of the highest Fock state included in the simulation (currently $\ket{2}$)
during the optimization, thus obtaining control sequences which avoid exciting the higher-lying states.
To verify the results, we finally simulate the optimized control sequence
using a higher truncation dimension ($4$ instead of~$3$).
The evolution does not change significantly for any of our sequences thus justifying our optimization method.

\subsection{Fock state optimization}

\begin{table*}[t]
\setlength{\tabcolsep}{0.8em}
\begin{tabular}{lccccc}
  \hline \hline
  parameter set & sequence type & dim & fidelity~$F$ & Wigner negativity~$M$ & figure\\
  \hline \hline
  \lecocqII{} & optimal control & 3 & 0.5699 & 0.0157
  \\
  && 4 & 0.5687 & 0.0155
  & \ref{fig:lecocq2_best_d4}\\
  \cline{2-6}
  & $\pi$-pulse & 3 & 0.5030 & 0\\
  && 4 & 0.5028 & 0 & \ref{fig:lecocq2_pi_d4}
  \\
  \hline
  \lecocqIII{}  & optimal control & 3 & 0.6021 & 0.0319
  \\
  && 4 & 0.5961 & 0.0303
  & \ref{fig:lecocq3_best_d4}\\
  \cline{2-6}
  & $\pi$-pulse & 3 & 0.5230 & 0.0007\\
  && 4 & 0.5230 & 0.0007 & \ref{fig:lecocq3_pi_d4}
  \\
  \hline \hline
\end{tabular}
\caption{
\label{table:results_fock}
Summary of Fock state optimization results.
The target state $\ket{\psi_T} = \ket{1}$ is a Fock state of the mechanical oscillator.
dim denotes the truncation dimension of the Hilbert spaces of the cavity and the oscillator used in the simulation.
The fidelity between a mixed final state~$\rho$ and a pure target state~$\ket{\psi_T}$
is $F(\rho, \ketbra{\psi_T}{\psi_T}) = \bra{\psi_T} \rho \ket{\psi_T}$,
which in this case is equal to the state $\ket{1}$ population of the oscillator.
As measure of the negativity of the Wigner function $W_\rho(\alpha)$ quantifying the non-classicality of the state $\rho$, 
we follow Ref.~\cite{albarelli_2018} and use
the so-called CV-mana $M(\rho) = \log \int \wrt{\alpha} |W_\rho(\alpha)|$.
}
\end{table*}

\begin{figure} 
  \hspace{5mm}(a)\hspace{0.5\columnwidth}(b)$\hfill$\\
  \includegraphics[width=0.48\columnwidth]{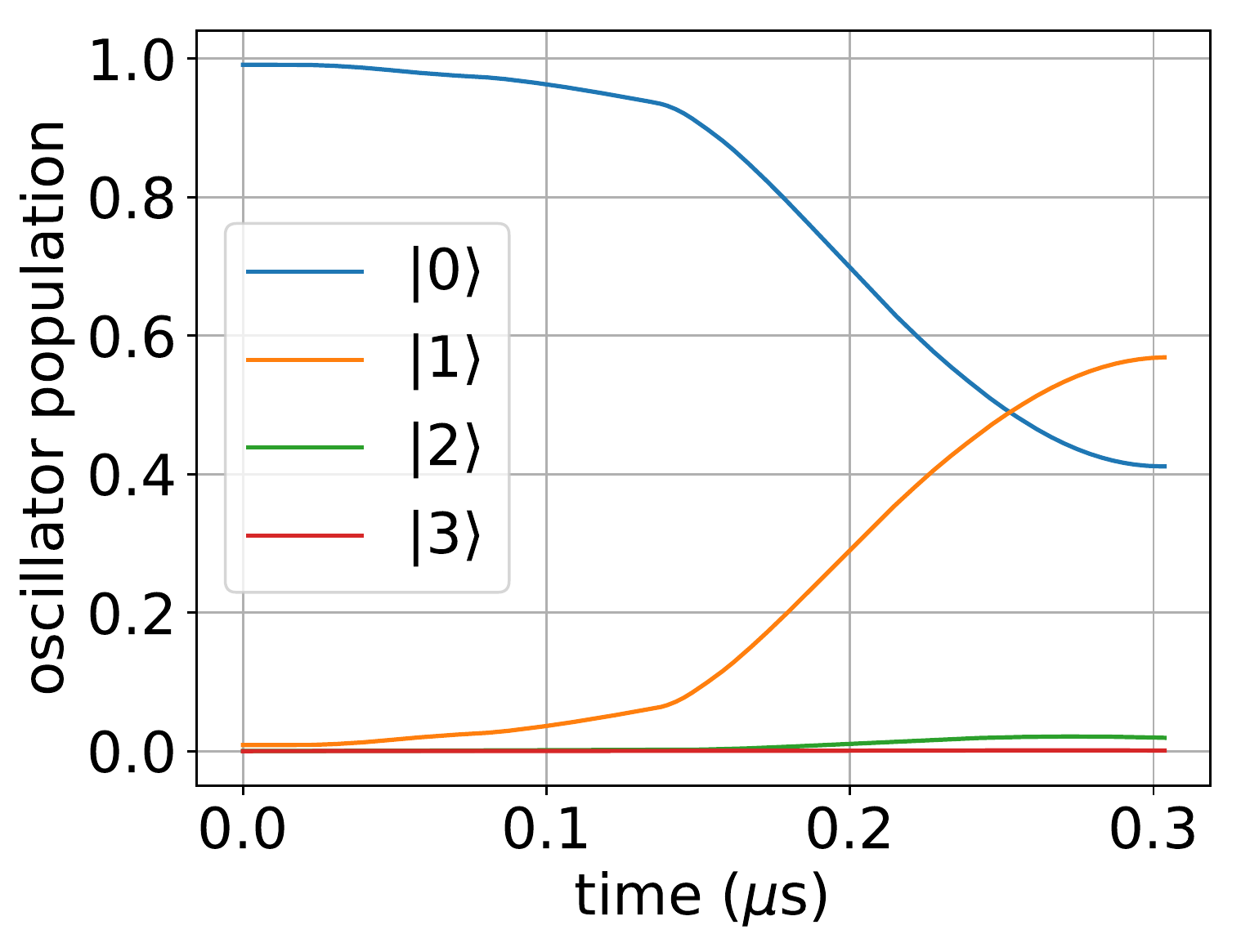}
  \includegraphics[width=0.48\columnwidth]{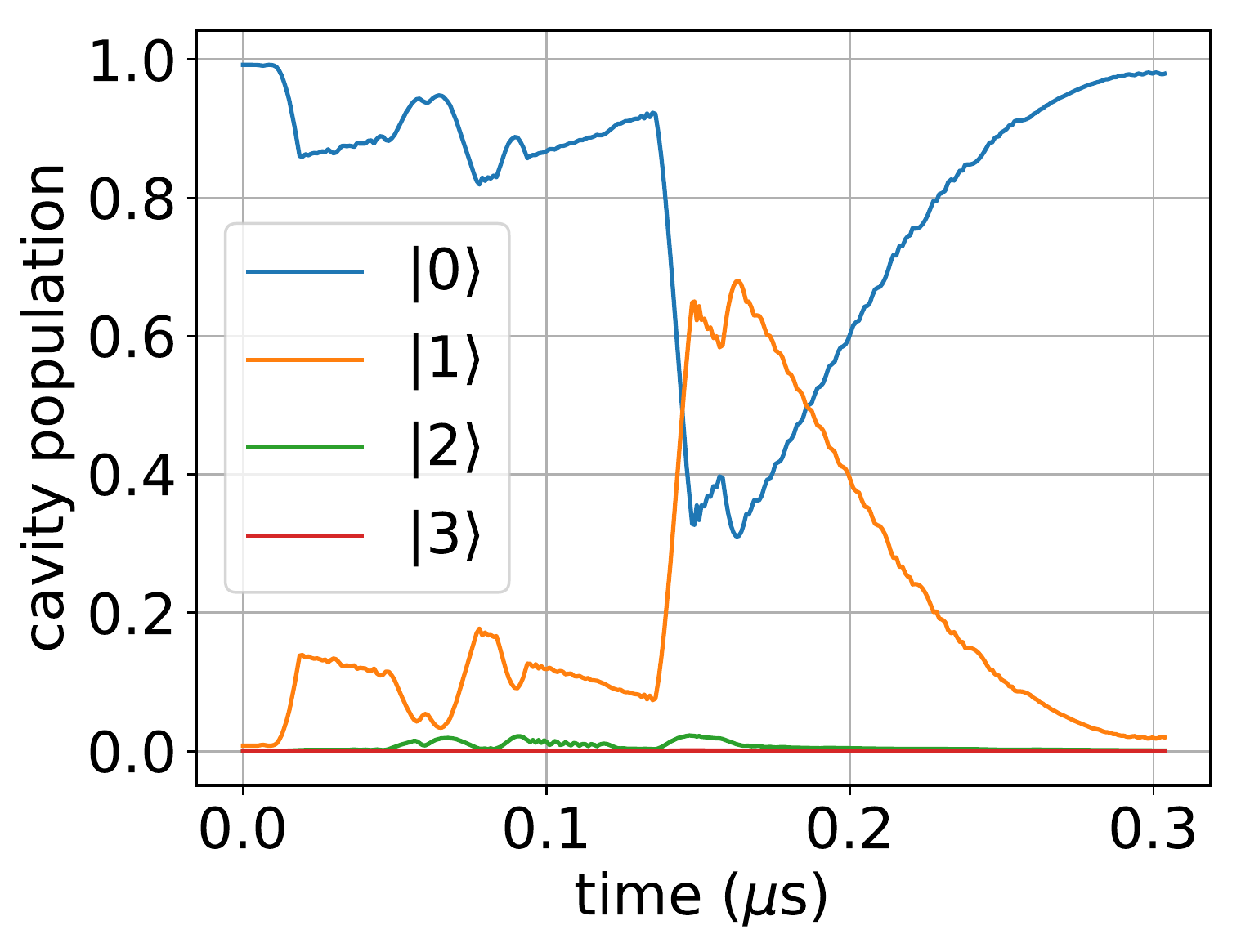}\\
  \hspace{5mm}(c)\hspace{0.5\columnwidth}(d)$\hfill$\\
  \includegraphics[width=0.44\columnwidth]{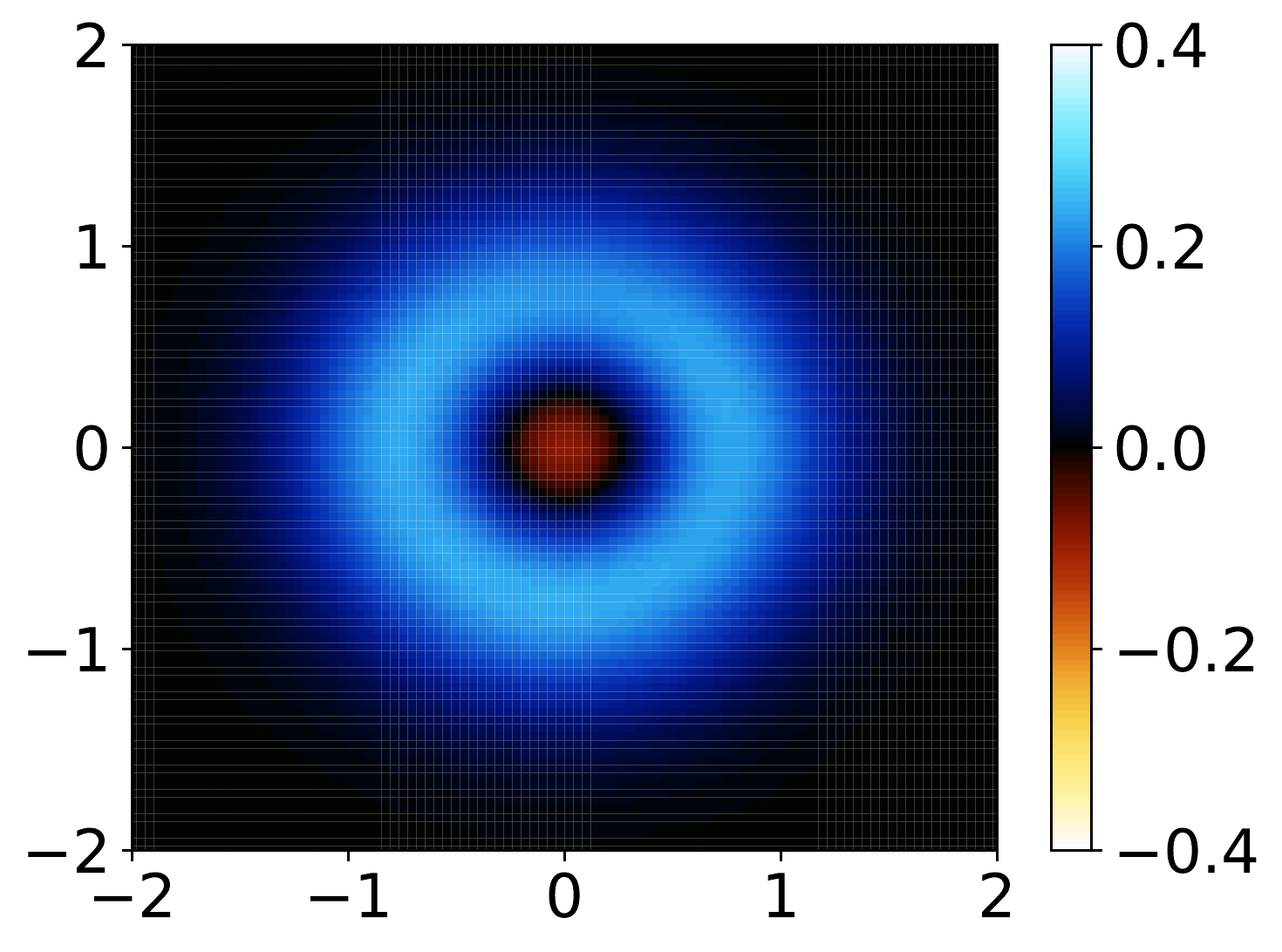}
  \includegraphics[width=0.52\columnwidth]{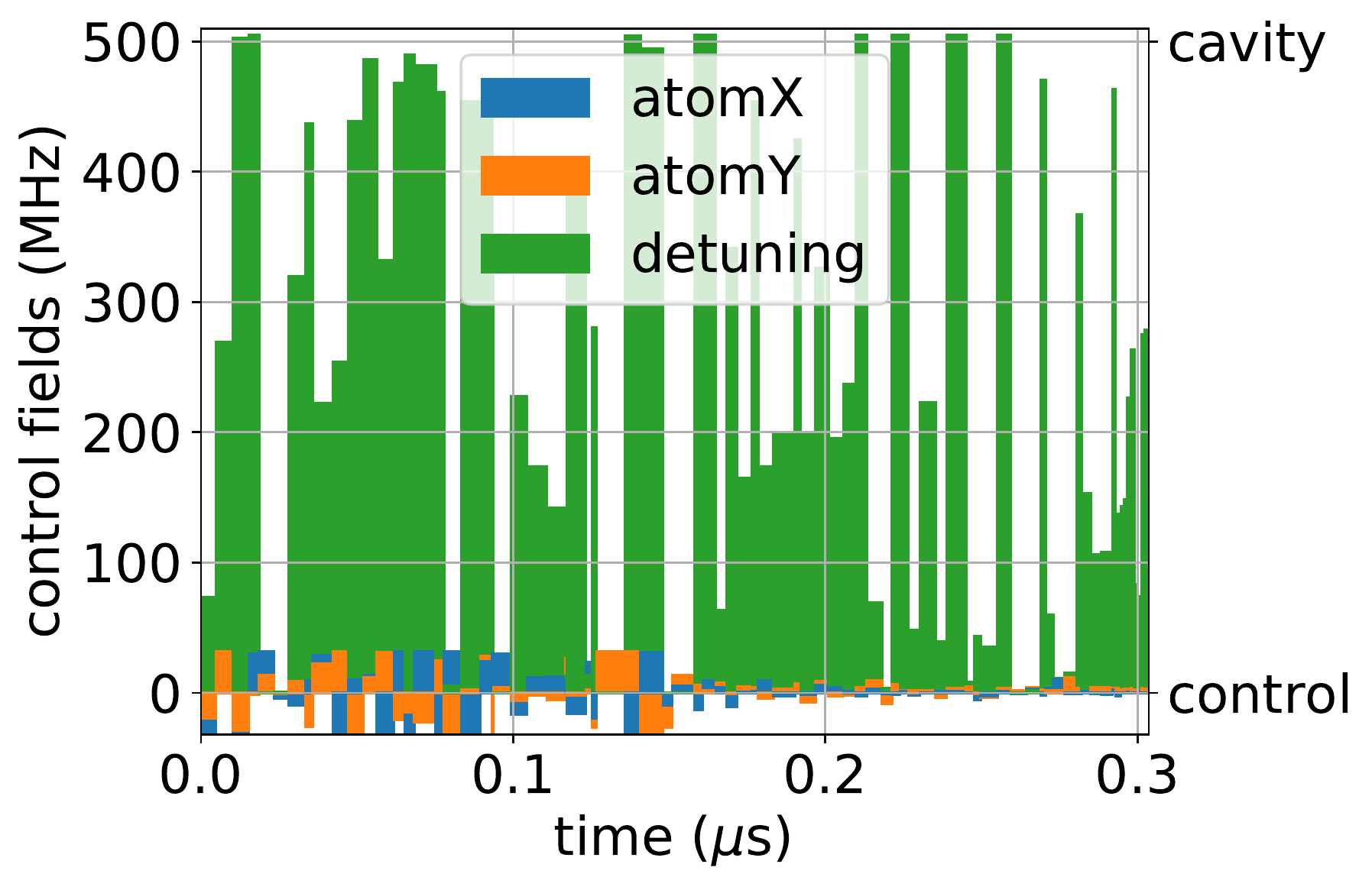}
  \caption{\label{fig:lecocq2_best_d4}
    Result of the Fock state $\ket{1}$ optimization using \lecocqII{} parameters.
    We can see that the states $\ket{2}$ and $\ket{3}$ are only slightly excited due to the penalty functional applied during the optimization.
    (a) oscillator population, (b) cavity population, (c) Wigner function of the oscillator at the end of the sequence, (d) optimized control sequence.
  }
\end{figure}

\begin{figure} 
  \hspace{5mm}(a)\hspace{0.5\columnwidth}(b)$\hfill$\\
  \includegraphics[width=0.48\columnwidth]{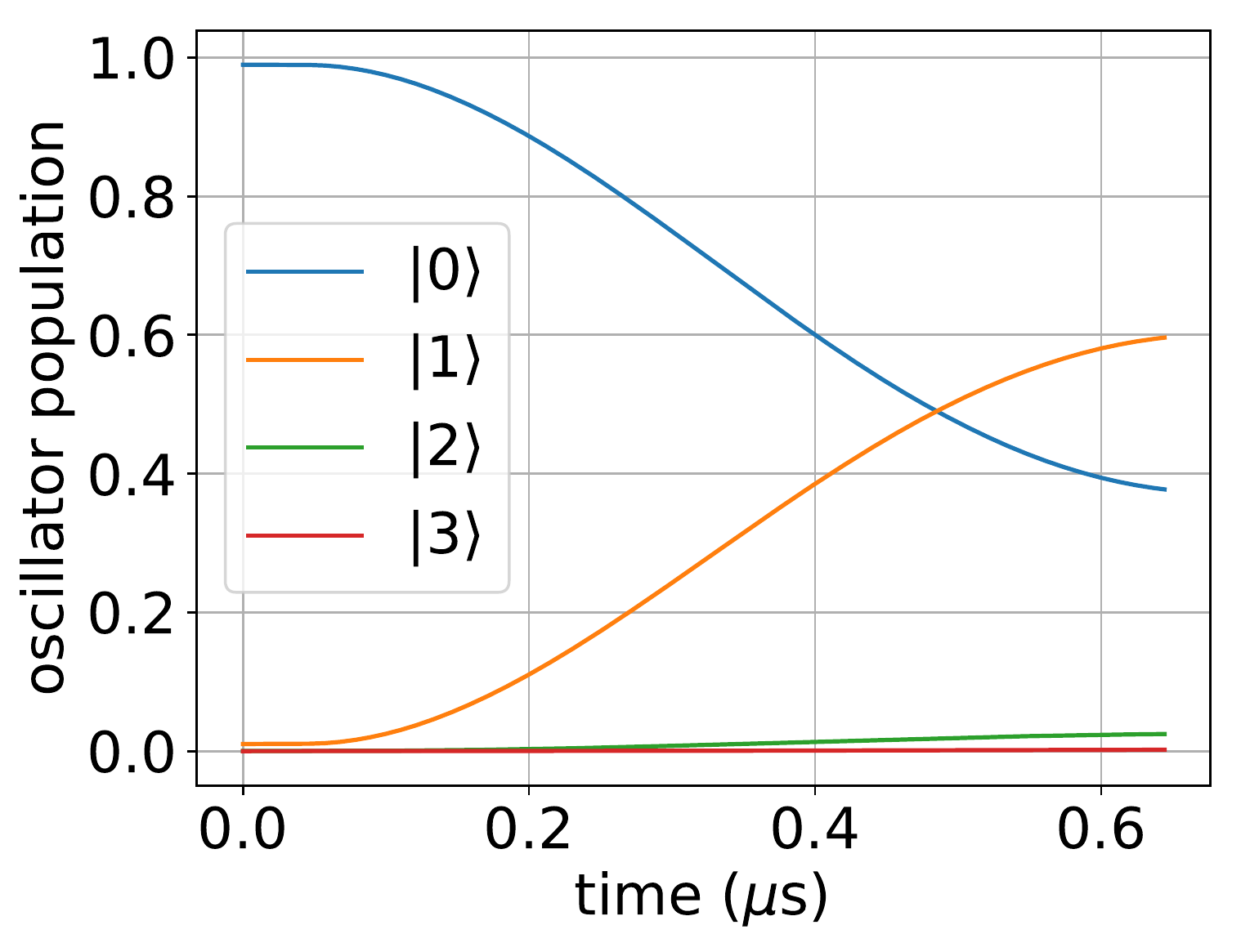}
  \includegraphics[width=0.48\columnwidth]{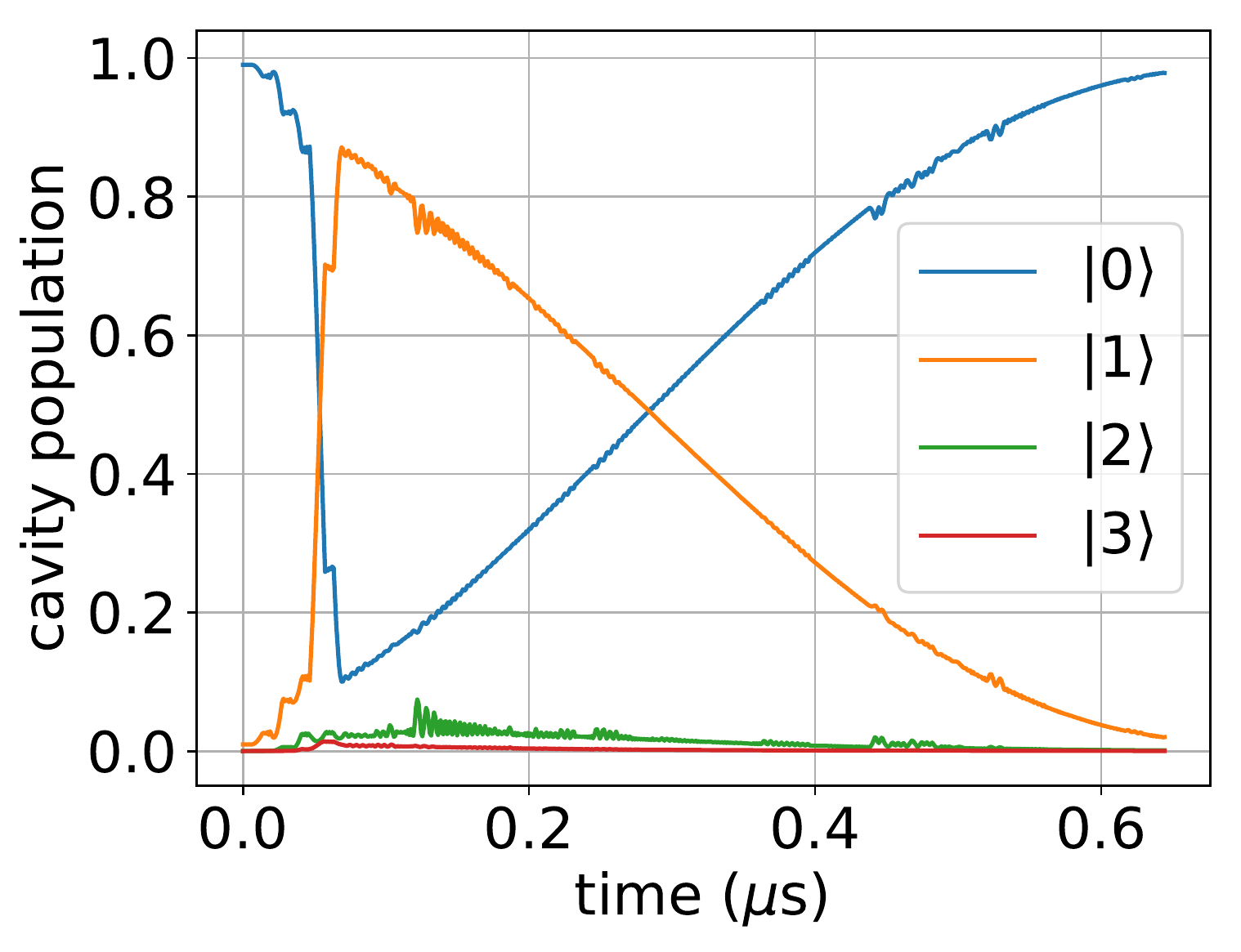}\\
  \hspace{5mm}(c)\hspace{0.5\columnwidth}(d)$\hfill$\\
  \includegraphics[width=0.44\columnwidth]{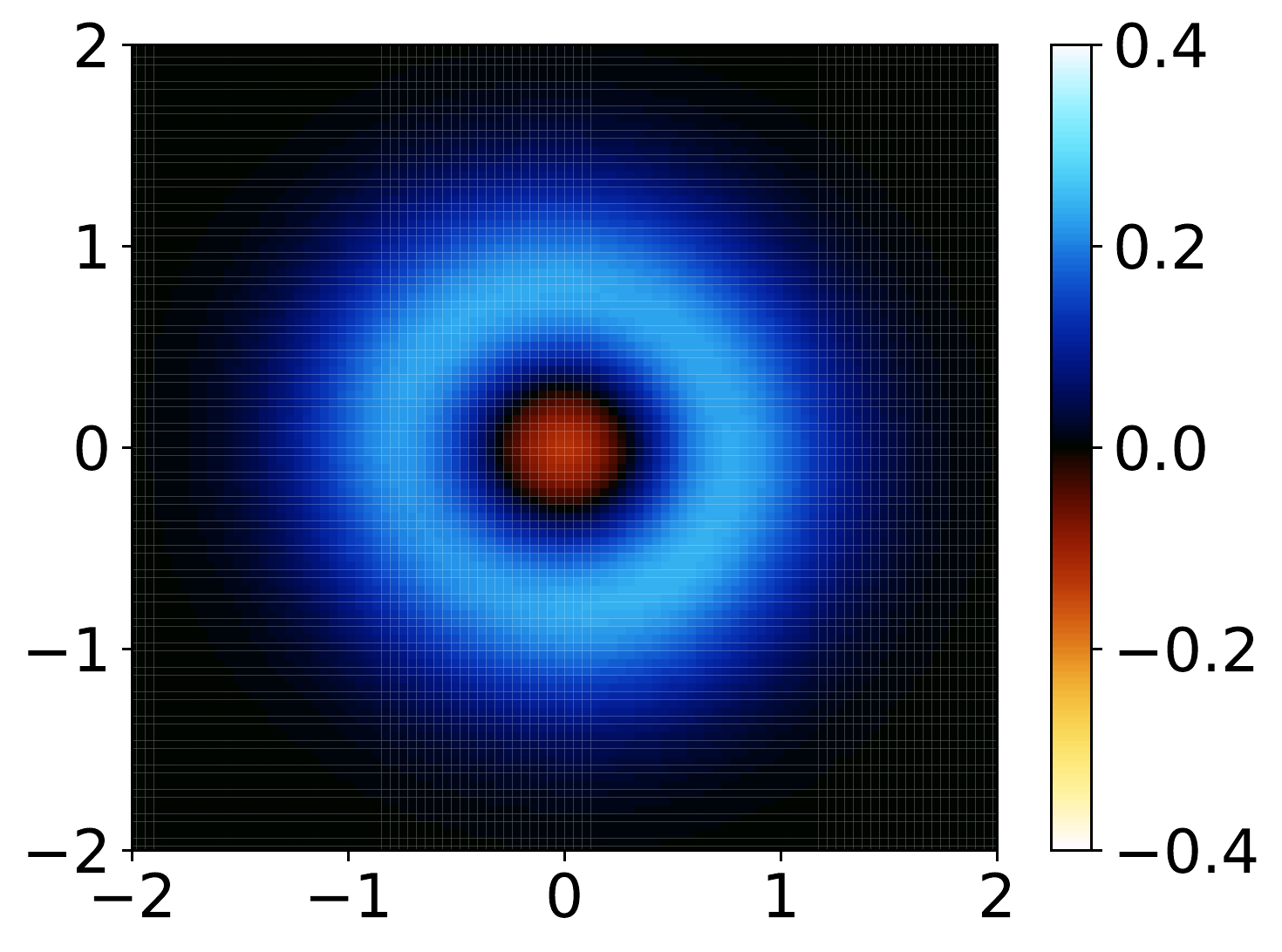}
  \includegraphics[width=0.52\columnwidth]{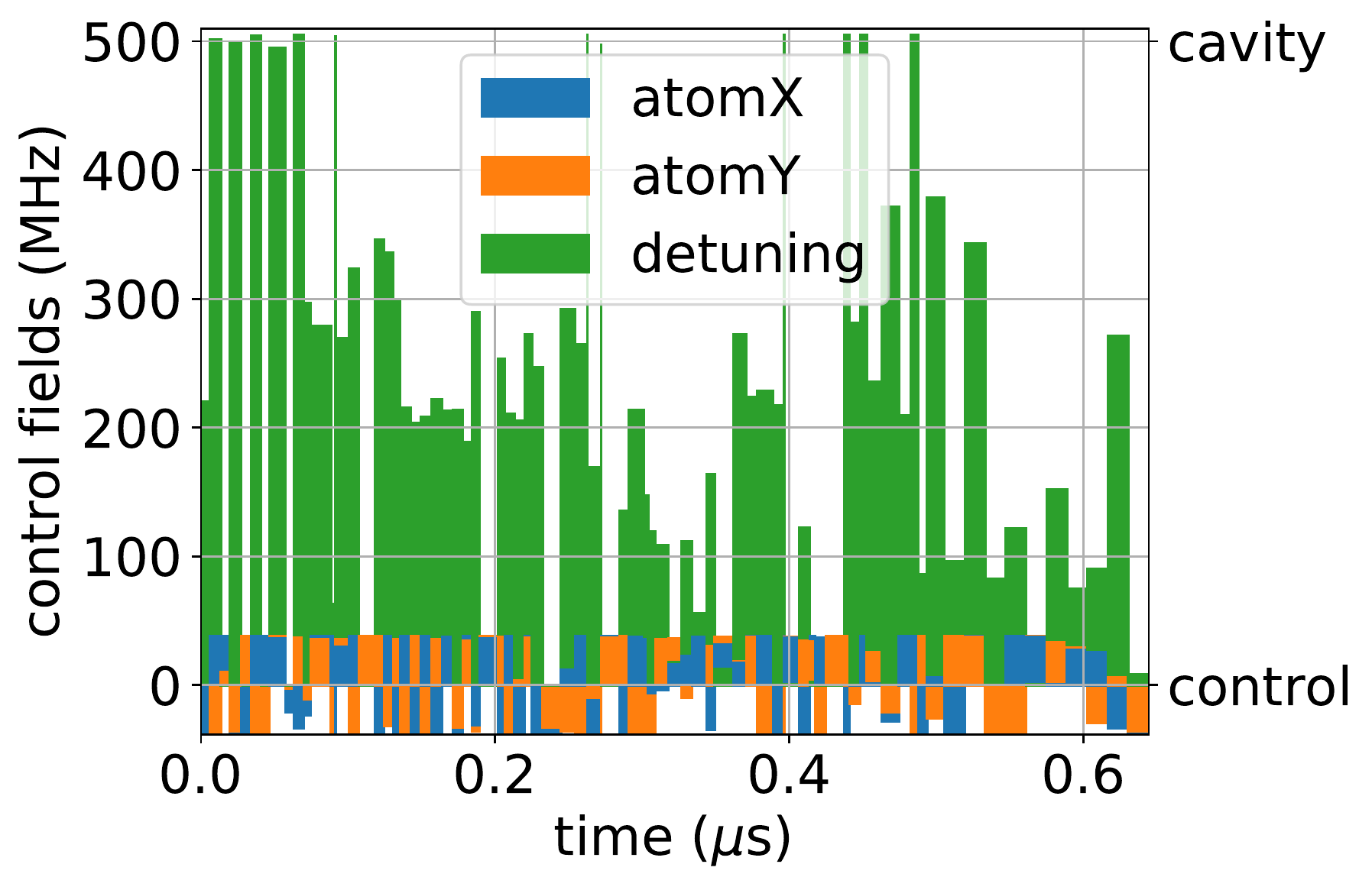}
  \caption{\label{fig:lecocq3_best_d4}
    Result of the Fock state $\ket{1}$ optimization using \lecocqIII{} parameters.
    Note that the states $\ket{2}$ and $\ket{3}$ are only slightly excited due to the penalty functional applied during the optimization.
    (a) oscillator population, (b) cavity population, (c) Wigner function of the oscillator at the end of the sequence, (d) optimized control sequence.
  }
\end{figure}

Here our optimization task is,
starting from the steady state of the system,
to create the Fock state~$\ket{1}$ in the oscillator,
without exciting the states~$\ket{2}$ and up
in either the cavity or the oscillator.
Note that the task cannot be accomplished exactly due to dissipation.

We quantify the non-classicality of the resulting oscillator state using
the CV-mana~\cite{albarelli_2018},
an easily computable monotone,
as the measure of Wigner negativity.
It is defined as the logarithm of the integral of the absolute value of the Wigner function,
$M(\rho) = \log \int \wrt{\alpha} |W_\rho(\alpha)|$.
It has the value zero for all classical states (i.e. states with nonnegative Wigner functions).
For the exact target state $\ket{1}$ we obtain $M(\ket{1}) \approx 0.355$.  
The purpose of the non-classicality measure, given a measurement procedure with a specific level of uncertainty,
is to say whether the measurement results expected in our state could have been produced by a classical state instead.

The results of the Fock state optimization are presented in Figs.~\ref{fig:lecocq2_best_d4} and~\ref{fig:lecocq3_best_d4},
and summarized in Table~\ref{table:results_fock}.
We notice that with both parameter sets we are able to obtain a clearly non-classical state (with the $\ket{1}$ population
significantly higher than the $\ket{0}$ population and the CV-mana noticeably larger than zero),
while keeping the excitation of the higher-lying states in both
the cavity and the oscillator to a minimum. As expected, \lecocqIII{}
yields a slightly better result. In both regimes, mere $\pi$ pulses
leave the system in a classical state or indistinguishably close to one,
while  optimal-control derived sequences attain significantly
non-classical states with fidelities being limited mostly by
dissipation.

\subsubsection*{Comparison to $\pi$ pulse sequences}

We may compare the optimized control sequences preparing the $\ket{1}$ Fock state in the oscillator
to a naive excitation transfer
control sequence consisting of just $\pi$~pulses (or their drift Hamiltonian analogs).
The sequence consists of three segments.
The first pulse excites the atom, the second moves the atom in resonance with the cavity until the excitation is transferred there,
and the final segment moves the atom back out of resonance and waits until the excitation has hopped from the cavity into the oscillator.
Since there are several simultaneously active interaction terms as
well as dissipation, this somewhat naive sequence does not perform very well.

\begin{figure}[ht!]
  \hspace{5mm}(a)\hspace{0.5\columnwidth}(b)$\hfill$\\
  \includegraphics[width=0.48\columnwidth]{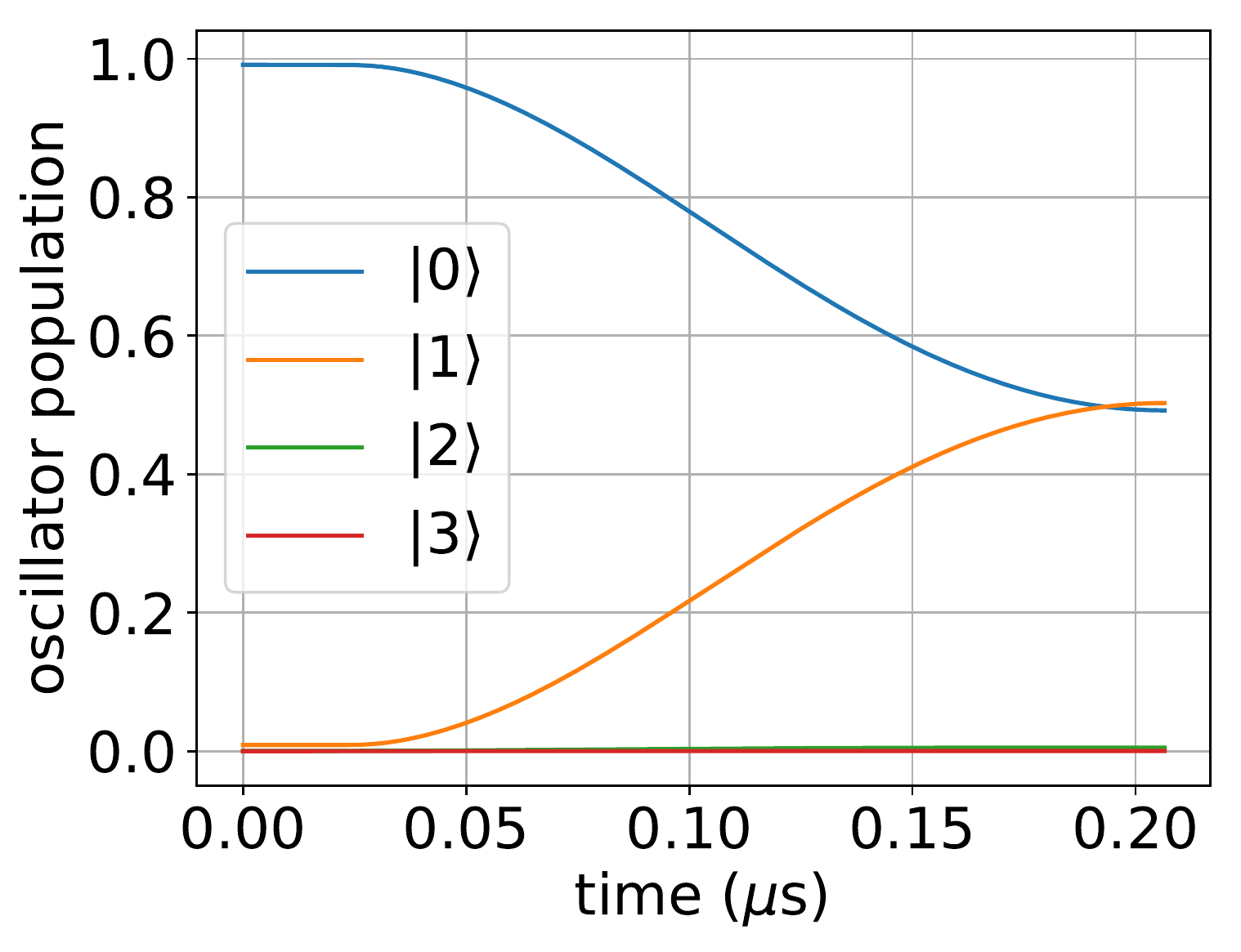}
  \includegraphics[width=0.48\columnwidth]{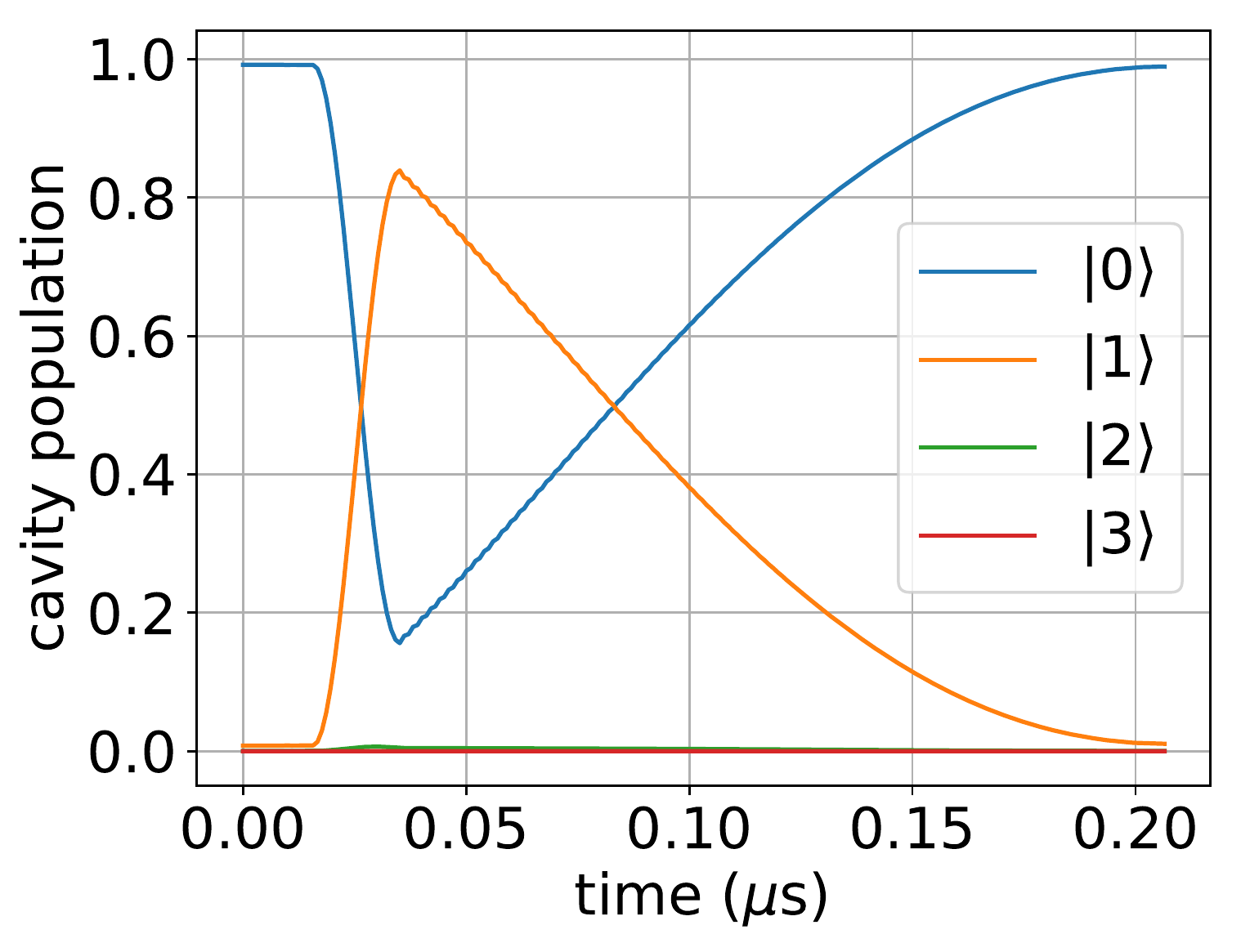}\\
  \hspace{5mm}(c)\hspace{0.5\columnwidth}(d)$\hfill$\\
  \includegraphics[width=0.44\columnwidth]{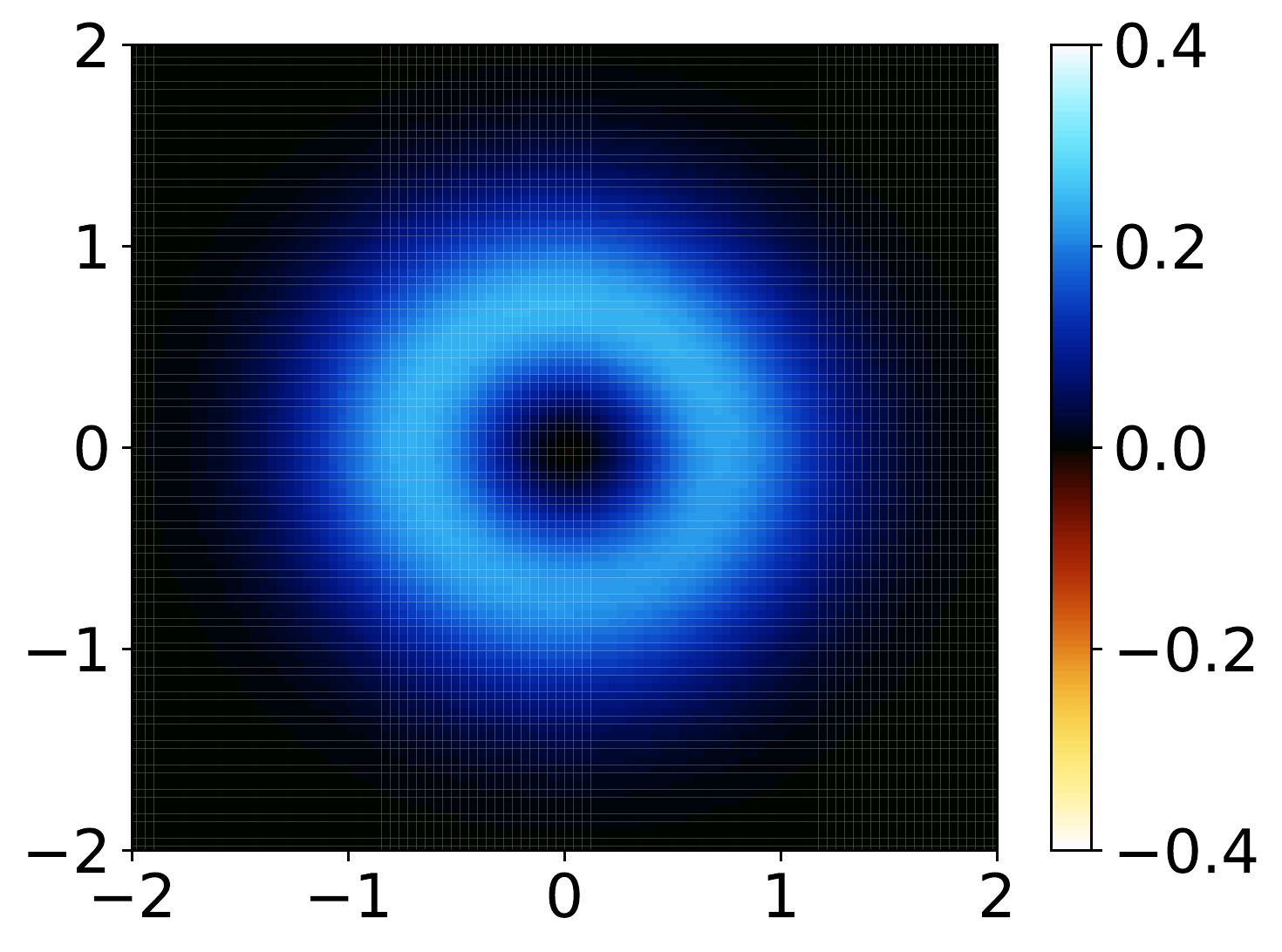}
  \includegraphics[width=0.52\columnwidth]{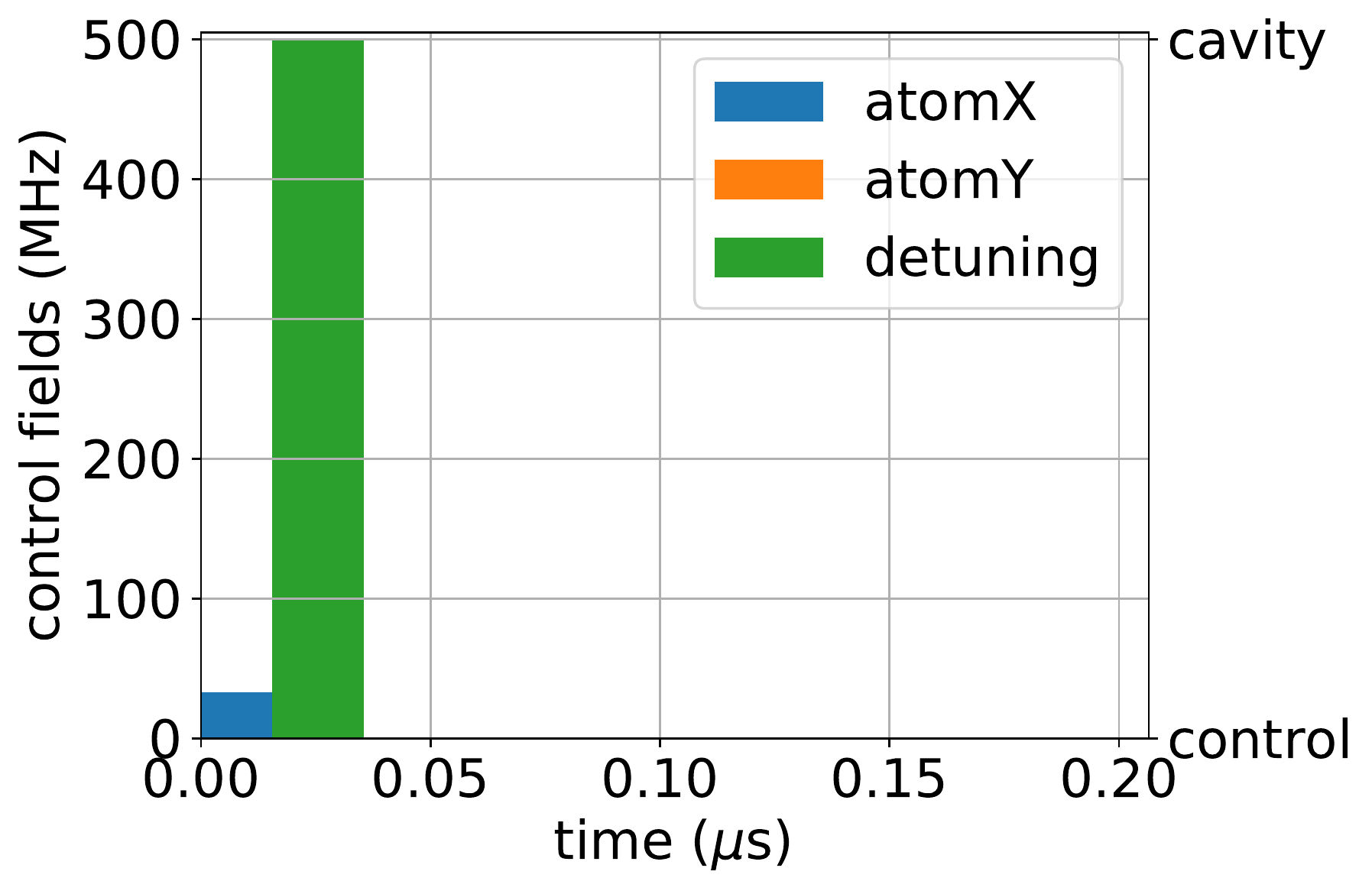}
  \caption{\label{fig:lecocq2_pi_d4}
    \lecocqII, excitation transfer using optimized $\pi$~pulses.
    (a) oscillator population, (b) cavity population, (c) Wigner function of the oscillator at the end of the sequence, (d) optimized control sequence.
  }
\end{figure}

\begin{figure}[ht!]
  \hspace{5mm}(a)\hspace{0.5\columnwidth}(b)$\hfill$\\
  \includegraphics[width=0.48\columnwidth]{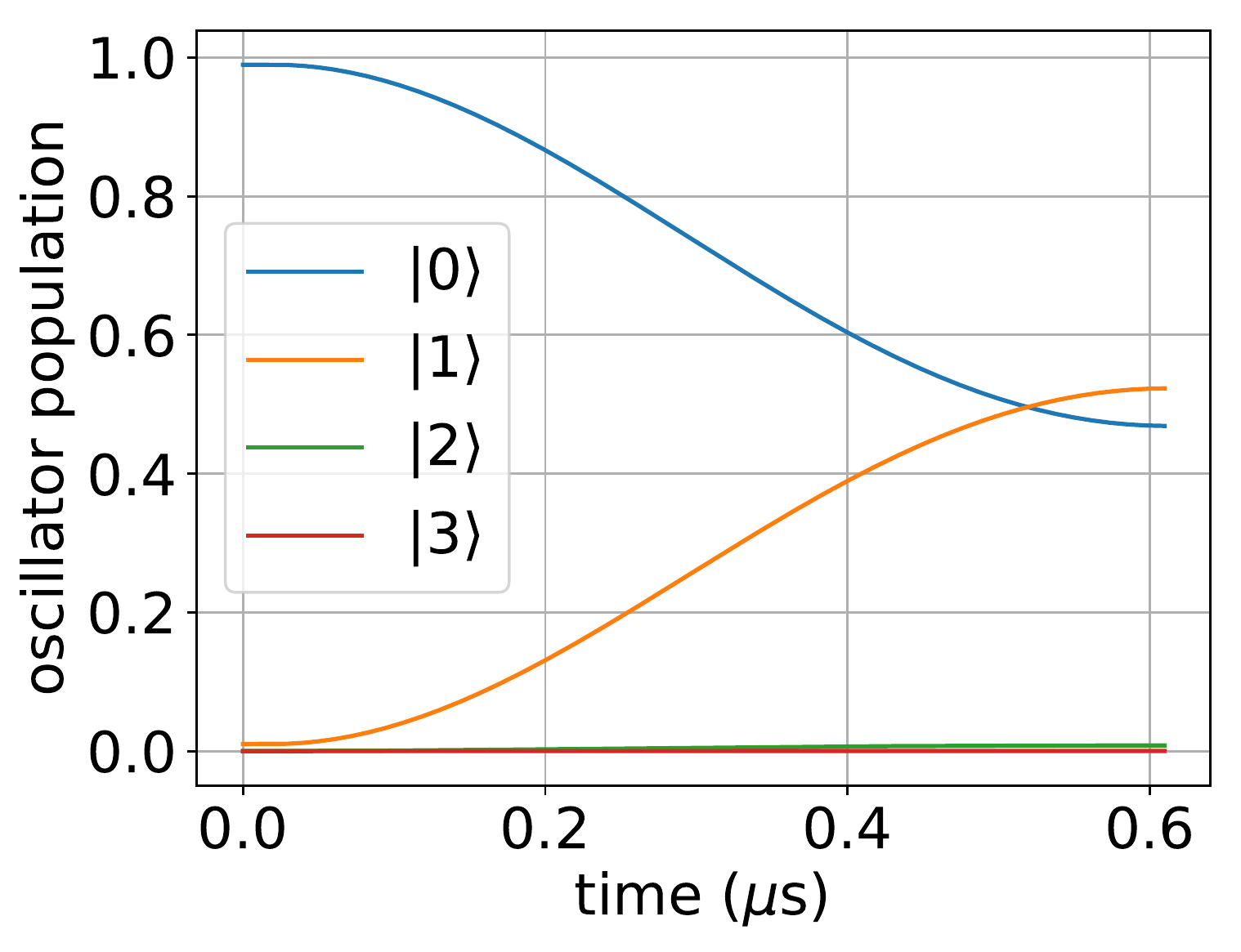}
  \includegraphics[width=0.48\columnwidth]{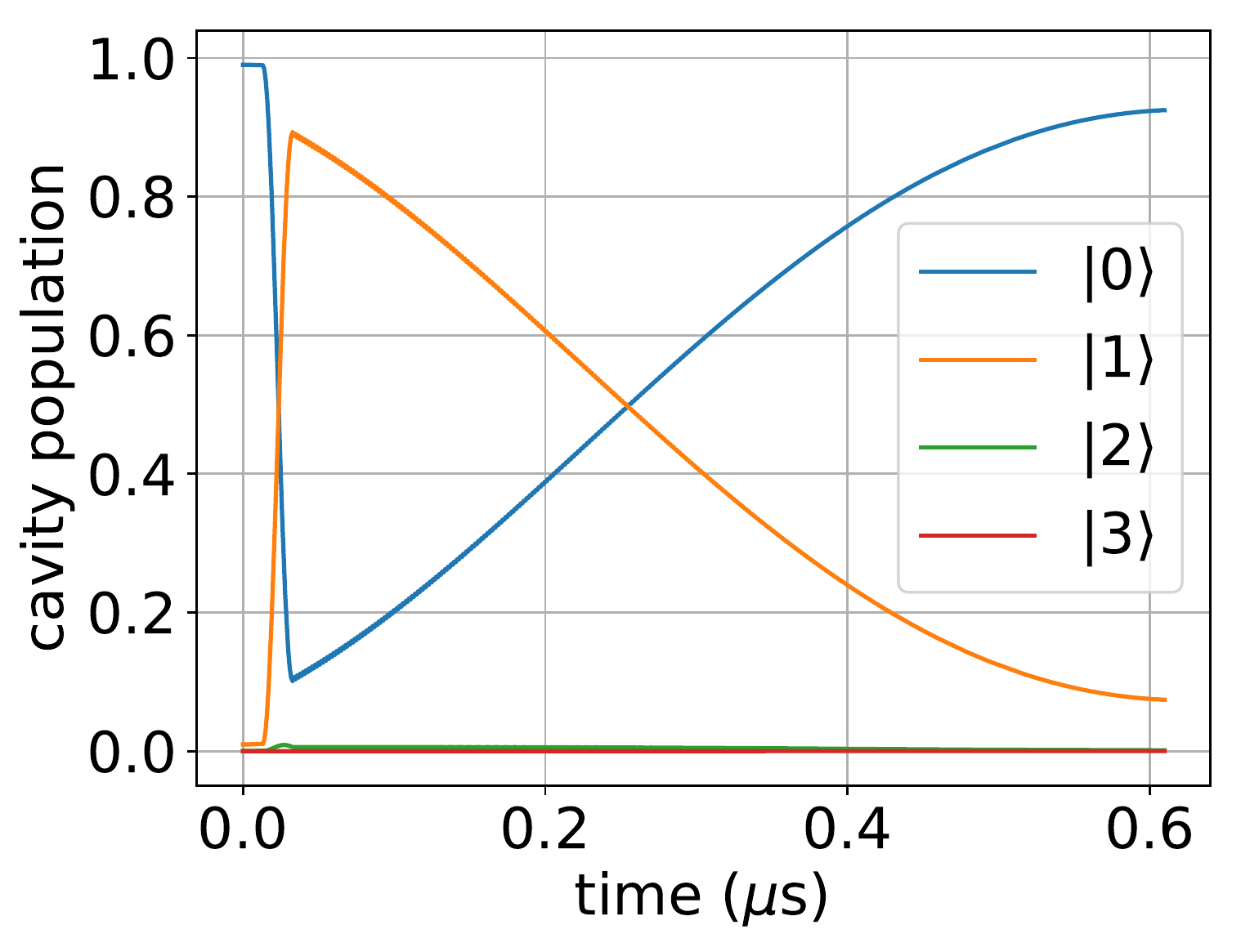}\\
  \hspace{5mm}(c)\hspace{0.5\columnwidth}(d)$\hfill$\\
  \includegraphics[width=0.44\columnwidth]{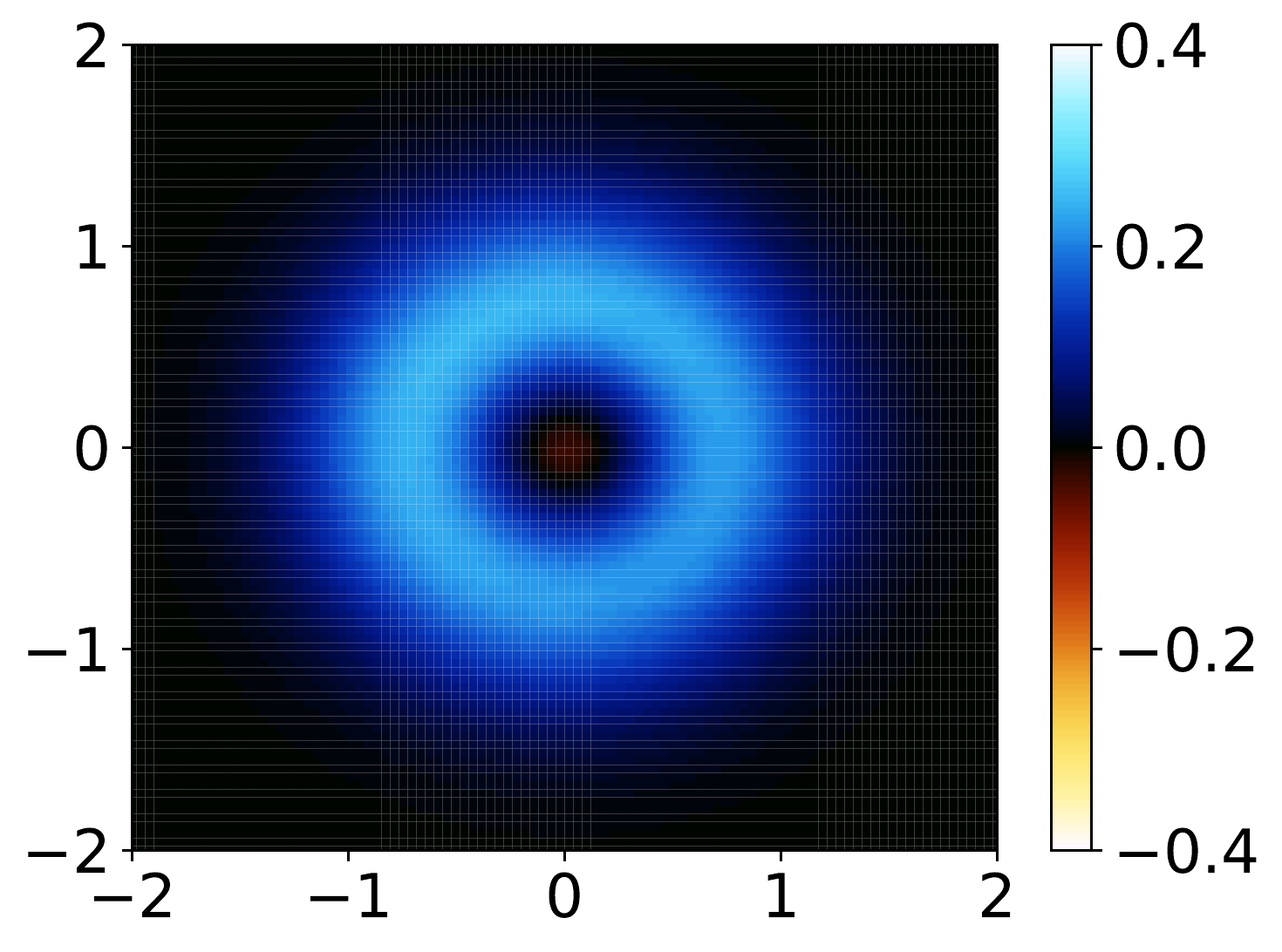}
  \includegraphics[width=0.52\columnwidth]{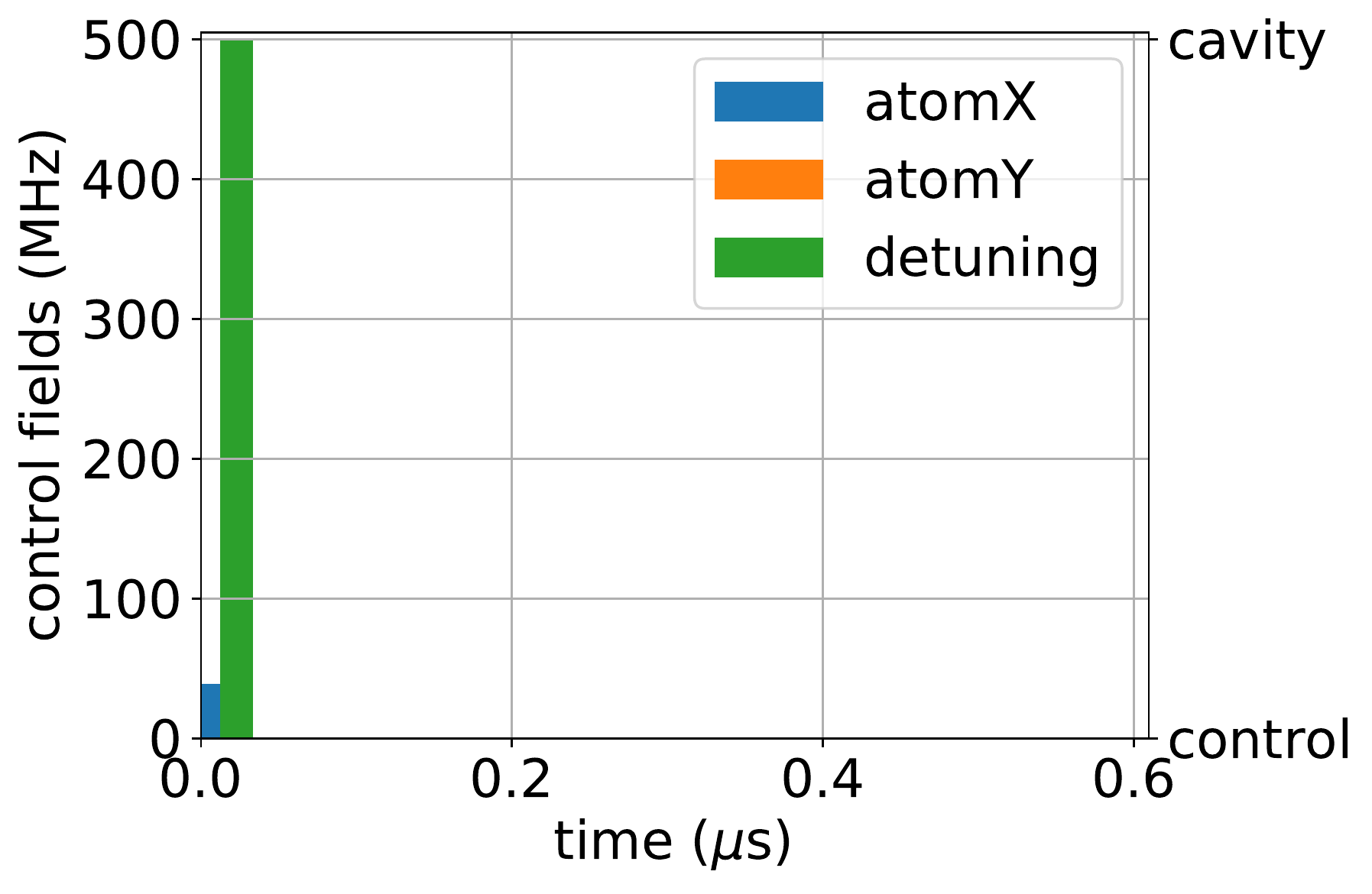}
  \caption{\label{fig:lecocq3_pi_d4}
    \lecocqIII, excitation transfer using optimized $\pi$~pulses.
    (a) oscillator population, (b) cavity population, (c) Wigner function of the oscillator at the end of the sequence, (d) optimized control sequence.
  }
\end{figure}

We may improve on it by optimizing the durations of each of the three pulses such that the population transfer during each step is maximized.
The optimized $\pi$-pulse sequences are presented in Figs.~\ref{fig:lecocq2_pi_d4} and~\ref{fig:lecocq3_pi_d4},
and their performance is summarized in Table~\ref{table:results_fock}.
Unlike the fully optimized sequences, the $\pi$ pulses facilitate observing the timescales of various interaction processes.
For example, in Fig.~\ref{fig:lecocq2_pi_d4}(b) the population of the $\ket{1}$ state of the cavity first go up from zero to
$0.84$ on the timescale $\tau=2\pi/(4g)$ of the atom-cavity interaction~$\gac/(2\pi) = 12.5$~MHz, and fall back to zero roughly on the timescale of the
boosted cavity-oscillator interaction~$\gco s/(2\pi) = 1.2$~MHz, expedited by dissipation.

With the \lecocqII{} parameters the $\pi$-pulse sequence fails to produce a substantially non-classical state, as can
be seen from the Wigner function which has a barely visible negative region in the middle.
With the \lecocqIII{} parameters the $\pi$ pulses fare a little better, but remain inferior
to the fully optimized control sequence, as shown in Table~\ref{table:results_fock}.

\subsection{Entangled state optimization}

\begin{table*}[t]
\setlength{\tabcolsep}{0.8em}
\begin{tabular}{lccccc}
  \hline \hline
  parameter set & sequence type & dim & fidelity~$F$ & log-negativity~$L$ & figure\\
  \hline \hline
   \lecocqIII{} & optimal control & 3 & 0.6451 & 0.4643 \\
  && 4 & 0.6450 & 0.4680 & \ref{fig:lecocq3_noon_d4}
  \\
  \hline \hline
\end{tabular}
\caption{
\label{table:results_NOON}
Summary of entangled-state optimization results.
The target state $\ket{\psi_T} = (\ket{01}+\ket{10})/\sqrt{2}$ entangles the cavity and the mechanical oscillator.
dim denotes the truncation dimension of the Hilbert spaces of the cavity and the oscillator used in the simulation.
The entanglement between the two bosonic modes is quantified using the logarithmic negativity
$L(\rho) = \log_2 \|\rho^{\text{PT}}\|_{\text{tr}}$.
}
\end{table*}

\begin{figure}
  \hspace{5mm}(a)\hspace{0.5\columnwidth}(b)$\hfill$\\
  \includegraphics[width=0.48\columnwidth]{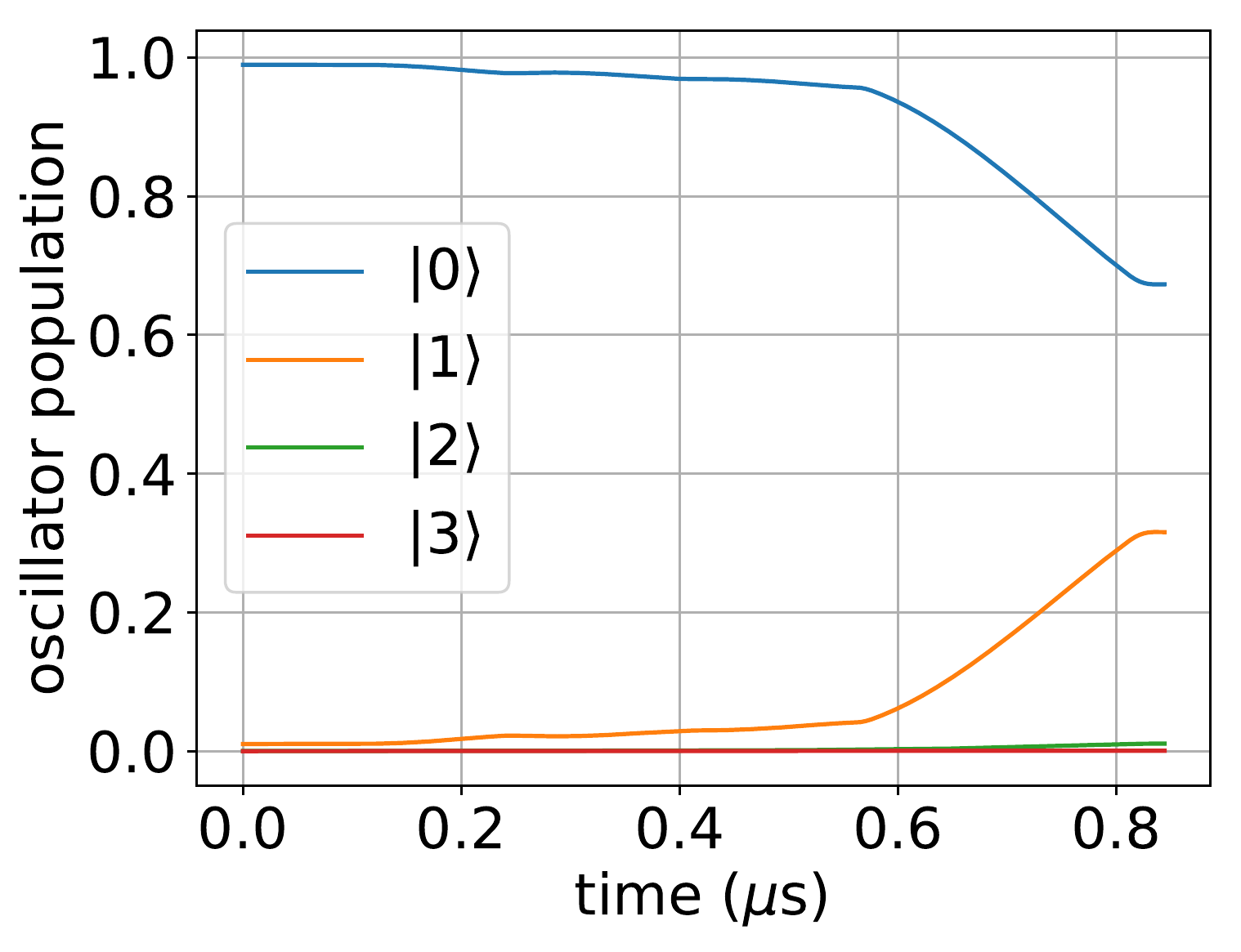}
  \includegraphics[width=0.48\columnwidth]{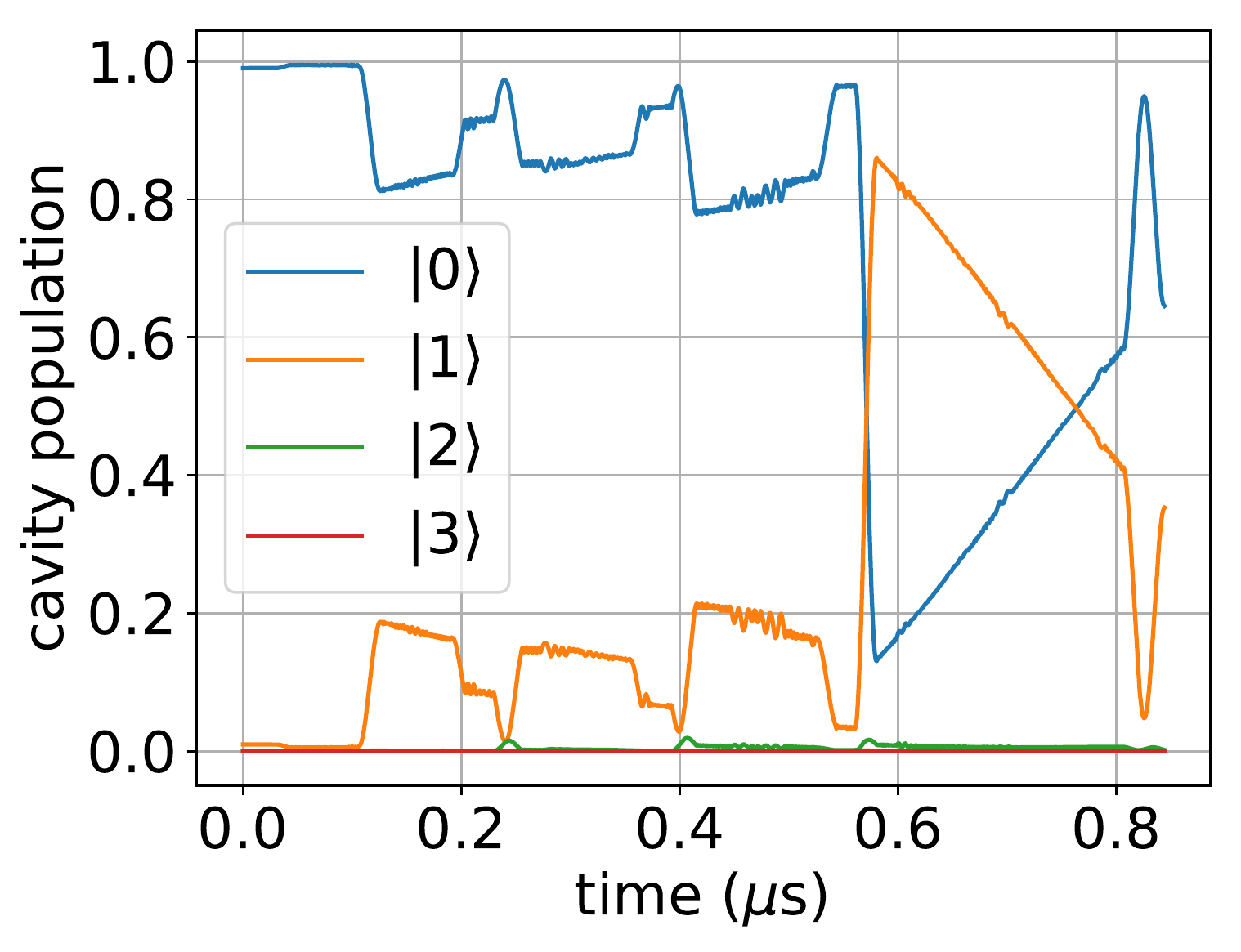}\\
  \hspace{5mm}(c)\hspace{0.5\columnwidth}(d)$\hfill$\\
  \includegraphics[width=0.48\columnwidth]{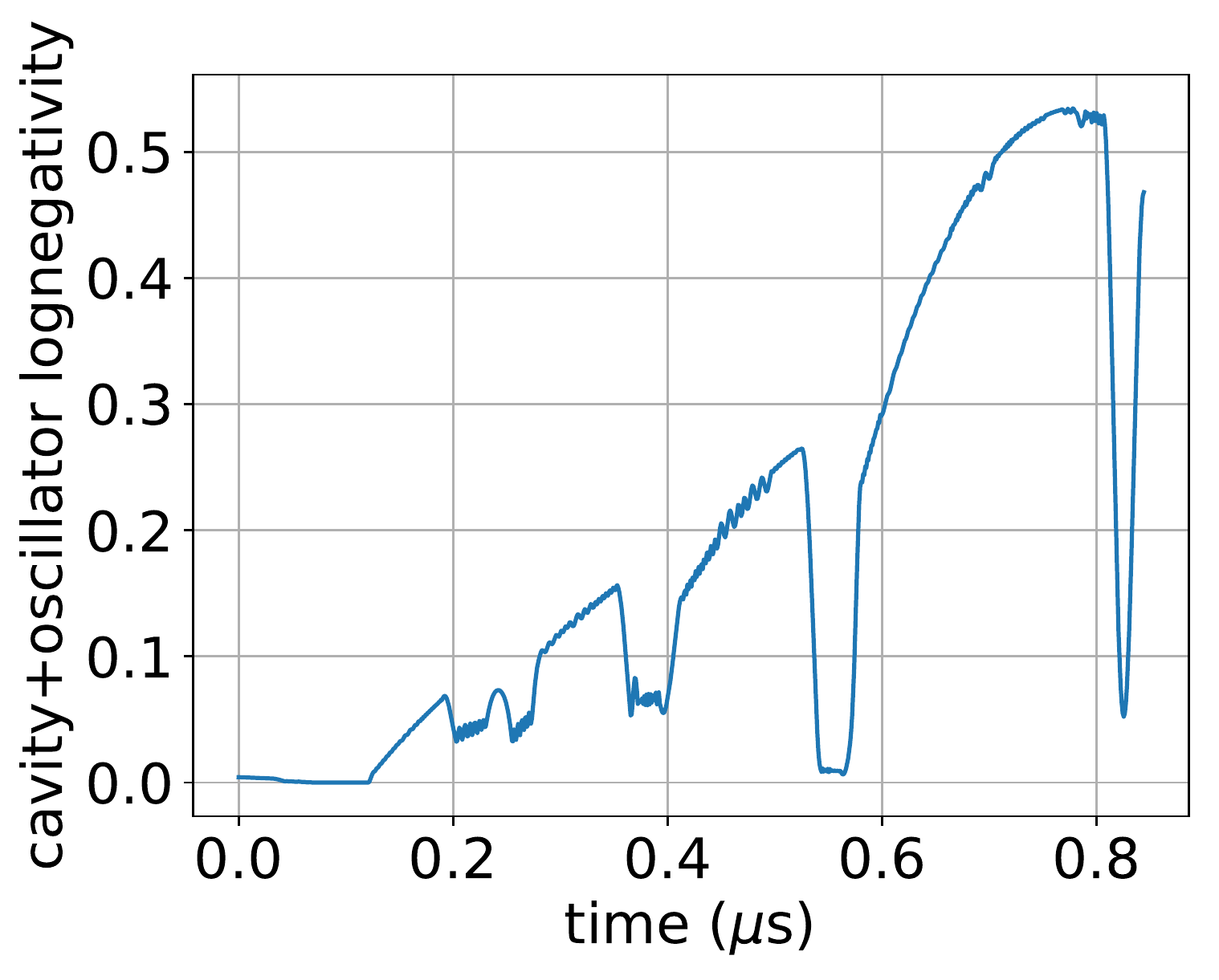}
  \includegraphics[width=0.48\columnwidth]{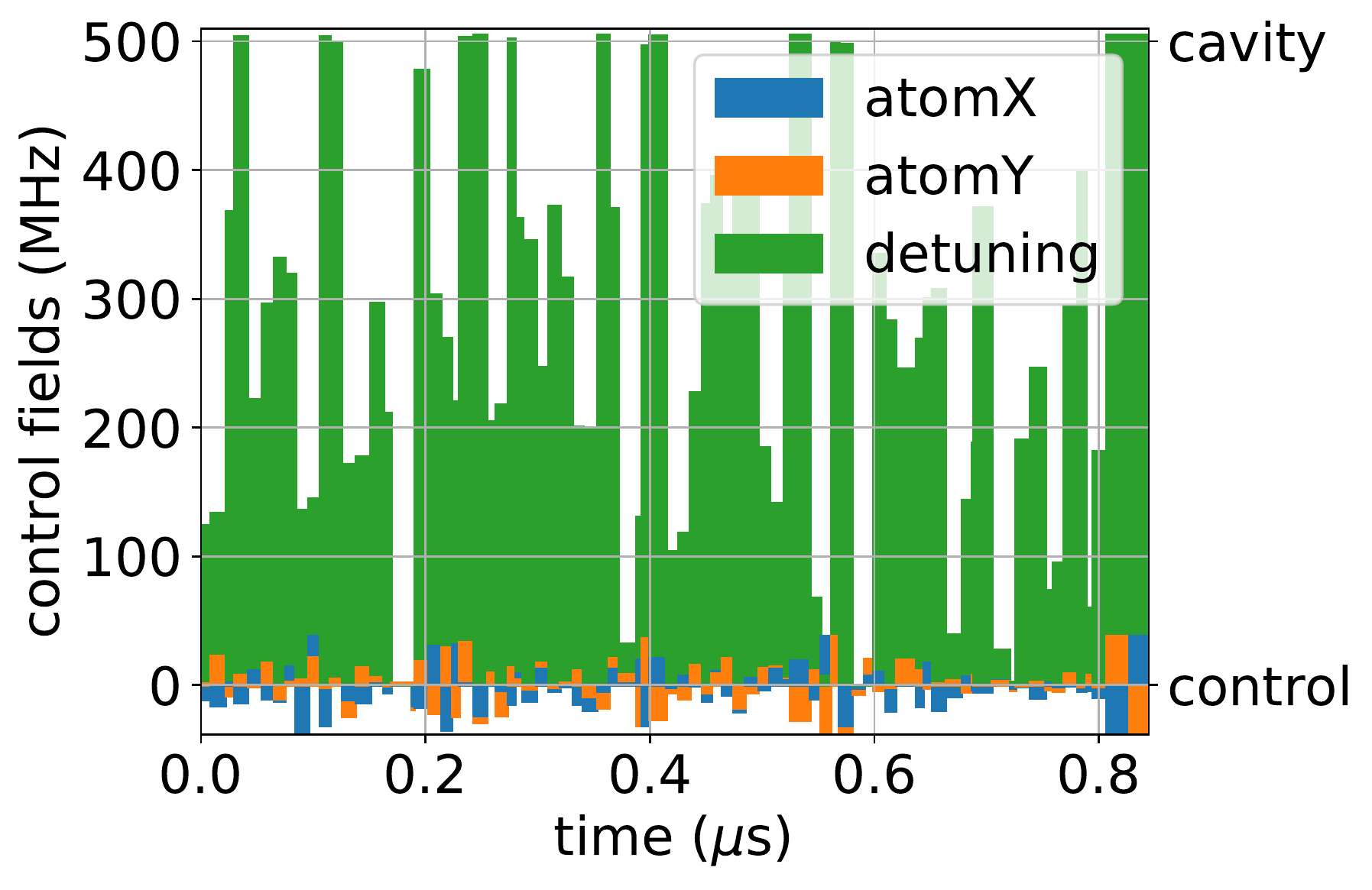}
  \caption{\label{fig:lecocq3_noon_d4}
    Result of the cavity-oscillator entangled state $(\ket{01}+\ket{10})/\sqrt{2}$ optimization using \lecocqIII{} parameters.
    The states $\ket{2}$ and $\ket{3}$ are only slightly excited due to the penalty functional applied during the optimization.
    (a) oscillator population, (b) cavity population, (c) logarithmic negativity of the cavity-oscillator state, (d) optimized control sequence.
  }
\end{figure}

Here we aim for a different target state, namely
the entangled cavity-oscillator state $(\ket{01}+\ket{10})/\sqrt{2}$.
Again, dissipation prevents us from achieving this exact state.
We quantify the entanglement between the optical and the mechanical mode using the logarithmic negativity
of the reduced cavity-oscillator state.
The logarithmic negativity of a bipartite state~$\rho$ is defined as
$L(\rho) = \log_2 \|\rho^{\text{PT}}\|_{\text{tr}} = \log_2(\sum_i s_i)$,
where $s_i$ are the singular values of the partial transpose of~$\rho$.
The logarithmic negativity is zero for all positive partial transpose (PPT) states (which include all separable states),
and has the value~$1$ for the exact target state.

The results of the entangled state optimization are presented in Fig.~\ref{fig:lecocq3_noon_d4},
and summarized in Table~\ref{table:results_NOON}.
With the \lecocqIII{} parameters
we are able to obtain a decidedly non-classical entangled optomechanical state
with minimal excitation of the higher-lying Fock states.


\section{Conclusions and Outlook}\label{sec:discussion}

In the present work we have shown how adding a steerable atom on top of a cavity coupled to a mechanical oscillator paves the way to 
(approximate) full controllability on the oscillator side. The system thus extended allows for preparing any state of the
harmonic oscillator subsystem from any initial state, within the limits imposed by dissipation.
More precisely, the extension overcomes the limitations of previous designs confined to a cavity coupled to an oscillator (without interaction to an atom), 
where linear feedback from homodyne detection came with the inevitable confinement to interconverting {\em within} equivalence classes of Gaussian oscillator states
or more generally of states with constant Wigner negativity. It is only by adding an interacting atom that
we obtain controlled dynamics including interconversion
{\em between} different equivalence classes of oscillator states.  

For illustration, we focused on generating the mechanical Fock state~$\ket{1}$,
and the optomechanical entangled state $(\ket{01}+\ket{10})/\sqrt{2}$,
truncating the control state space at dimension $d=3$. However, higher truncations at $d=5$ or larger are imaginable.
A larger control state space would allow studying the generation of further non-classical states of interest,
such as mechanical Schr\"odinger cat states~\cite{mancini_ponderomotive_1997,brunelli_unconditional_2018}
or higher NOON states~\cite{ren_single-photon_2013}, relevant for studying macroscopic non-classicality~\cite{marshall_towards_2003},
or cubic phase states~\cite{yukawa_emulating_2013,houhou_unconditional_2018}, relevant for Gaussian quantum computation~\cite{houhou_unconditional_2018}.

Another aspect, where optimal control may be important, is to
account for the multimode character of the mechanical oscillator
\cite{wieczorek_optimal_2015,nielsen_multimode_2017}. In particular,
when using pulsed control schemes, multiple mechanical modes lying in
the finite bandwidth of the pulsed optical drive will be addressed
simultaneously. This might lead to undesired optomechanical
correlations, which could readily be treated by including
Lagrange-type penalties into the target function subject to optimal
control.

Optimal control techniques giving non-classical mechanical states are thus anticipated to find future application, e.g., in nano-optic \cite{tiecke_nanophotonic_2014,chan_laser_2011}, ion-trap \cite{neumeier_reaching_2018} or circuit QED implementations \cite{lecocq_resolving_2015,schmidt_ultrawide-range_2018} of hybrid quantum optomechanical systems.

\begin{acknowledgments}
We acknowledge fruitful discussions with Giulia Ferrini and Markus Aspelmeyer and, at an earlier stage, Stephan Welte. We are grateful to an anonymous referee for highly constructive comments. The work was supported in part by the Excellence Network of Bavaria ({\sc enb}) through the programme {\em Exploring Quantum Matter}\/ (ExQM) and by {\em Deutsche Forschungsgemeinschaft} ({\sc dfg}, German Research Foundation) under Germany’s Excellence Strategy {\sc exc}-2111–390814868.
W.W.~acknowledges funding from Chalmers' Excellence Initiative Nano and from the \emph{Knut and Alice Wallenberg Foundation}.
\end{acknowledgments}

\bibliography{optomech}

\clearpage
\appendix
\onecolumngrid

\section{Deriving the optomechanical Hamiltonian}
\label{sec:deriv-optom-hamilt}

\subsection{Introduction}

The simplest realization of an optomechanical system is a single-mode Fabry-P\'erot cavity
with one mirror semitransparent for coupling to the outside world,
and the other mirror attached to a sufficiently harmonic mechanical oscillator~\cite{optomech2014}.
The optical cavity is driven through the semitransparent mirror using laser(s).
There are also other systems that follow similar dynamics,
e.g. quantum electromechanical circuits, and the discussion below applies to them as well.

Let us assume that there is just a single cavity mode and a single oscillator mode that are relevant.
We denote the annihilation operators of the cavity and the oscillator by~$\ha$ and $\hb$,
and the corresponding dimensionless position and momentum by~$(\hat{Q},\hat{P})$ and $(\hat{q},\hat{p})$, respectively.\footnote{
  We use $u_q u_p /\hbar = \frac{1}{2}$, where $u_q$ and $u_p$ are the units of the position and momentum quadratures,
  i.e. the dimensionless position and momentum operators of the cavity are
  $\hat{Q} = \hat{a}+\hat{a}^\dagger$ and $\hat{P}=-i(\hat{a}-\hat{a}^\dagger)$, and those of the oscillator are
  $\hat{q} = \hat{b}+\hat{b}^\dagger$ and $\hat{p}=-i(\hat{b}-\hat{b}^\dagger)$.}
Let the zero-point motion of the mechanical oscillator have the standard deviation~$x_0 = u_q \sqrt{\bra{0}\hat{q}^2\ket{0}}$.
The resonance frequency of the optical cavity~$\wc$ depends on its length, which is modulated by the position of the mechanical oscillator,
given by $\hat{x} = x_0 \hat{q} = x_0 (\hb+\hb^\dagger)$.
Linearizing, we obtain
\be
\wc(\hat{x}) \approx \wc -G \hat{x} = \wc -G x_0 (\hb+\hb^\dagger)
= \wc -\gco (\hb+\hb^\dagger).
\ee
In the lab frame the optomechanical system is thus described by the Hamiltonian
\be
\label{eq:om:co}
\Hco/\hbar = \wc \ha^\dagger \ha +\Omega_m \hb^\dagger \hb -\gco \ha^\dagger \ha (\hb+\hb^\dagger) +E(t) \cos(\wl t +\phi_L(t)) (\ha+\ha^\dagger),
\ee
where the last term represents driving of the optical cavity by a laser.
The $\hat{Q}$ quadrature of the cavity is defined as the direction of the driving.
The driving Rabi frequency is connected to the laser power~$P$ by
$E = \sqrt{\frac{2\kappa P}{\hbar \wl}}$.

The system may be made more controllable by adding a controllable two-level atom in the cavity,
with the Hamiltonian
\begin{align}
\label{eq:om:atom}
\Ha/\hbar
=
\wa \hat{\sigma}_+ \hat{\sigma}_-
+\gac (\ha e^{i\phi_c}  +\hc)(\hs e^{i\phi_a} +\hc)
+\ac \cos(\wr t +\phi_R(t))(\hs +\hc).
\end{align}
Above, the three terms represent the atom itself, the atom-cavity coupling,
and a classical control signal driving the atom, respectively.
The driving defines the atomic $X$~direction, and we (for now)
introduce the arbitrary phases $\phi_c$ and~$\phi_a$ to keep the
atom-cavity interaction term as generic as possible.

The dissipation processes in the cavity are described
by the Lindblad operator $\sqrt{\kappa} \ha$.
This assumes that the effective temperature of the cavity surroundings is zero,
which is an excellent approximation for microwave cavities cooled to ultra-low temperatures or for optical cavities operating at room temperature.
Likewise, the atomic decay is described by the Lindblad operator~$\sqrt{\kappa_a} \hat{\sigma}_-$.
For the mechanical oscillator, due to its lower resonance frequency, we need
both the annihilation and creation Lindblad operators
$\{\sqrt{\gamma'} \; \hb, \sqrt{\gamma' x} \; \hb^\dagger\}$, where $\gamma' = \gamma (\nn+1)$ is the effective decay rate,
$\nn = x/(1-x)$ is the expected number of oscillator phonons in the steady state given by the Bose-Einstein distribution function,
and $x = e^{-\frac{\hbar\Omega_m}{kT}}$ is the oscillator Boltzmann factor fulfilling $0 \le x < 1$.

The summary of the symbols used can be found in Table~\ref{table:symbols} along with the numerical values used in the simulations.

\subsection{Moving into a rotating frame}

To fix the terms driving the atom and the cavity we transform into a frame co-rotating with their frequencies,
$H_0/\hbar = \wl \ha^\dagger \ha +\wr \hat{\sigma}_+ \hat{\sigma}_-$,
obtaining
\begin{align}
\label{eq:om:drive}
\Hco'/\hbar =& \underbrace{(\wc-\wl)}_{-\Delta} \ha^\dagger \ha +\Omega_m \hb^\dagger \hb -\gco \ha^\dagger \ha (\hb+\hb^\dagger)
+\frac{E}{2}(\ha (e^{i\phi_L} +e^{-i(2\wl t +\phi_L)})  +\hc),
\end{align}
where $\Delta$ is the detuning between the laser and the cavity, and
\be
\begin{aligned}
\Ha'/\hbar
=\:&
(\wa-\wr) \hat{\sigma}_+ \hat{\sigma}_-
+\gac (\ha e^{i(-\wl t +\phi_c)}  +\hc)(\hs e^{i(\wr t+\phi_a)} +\hc)\\
&+\ac \cos(\wr t +\phi_R(t))(\hs e^{i\wr t} +\hc)\\
=\:&
(\wa-\wr) \hat{\sigma}_+ \hat{\sigma}_-
+\gac (\ha \hs e^{i((\wr-\wl) t +\phi_c+\phi_a)}
+\ha^\dagger \hs e^{i((\wl+\wr) t -\phi_c+\phi_a)} +\hc)\\
&+\frac{\ac}{2}(\hs(e^{-i\phi_R}+e^{i(2\wr t +\phi_R)}) +\hc)
\end{aligned}
\ee
We may then perform a rotating wave approximation and drop all three counter-rotating terms (and their hermitian conjugates).

The Lindblad operators in the rotating frame acquire a rotating complex phase factor which has no effect on the dynamics
since it cancels out.

\subsection{Shifting and rotating the cavity and oscillator states}
\label{sec:op_shift}

Ignoring the oscillator for the moment (setting $\gco = 0$),
with constant laser driving a pure coherent steady state~$\ket{\alpha}$ forms
in the cavity, where
\be
\label{eq:alpha}
\alpha = \frac{e^{-i\phi_L} E/2}{\Delta +i\kappa/2}.
\ee
If the average photon number $\expect{\ha^\dagger \ha} = |\alpha|^2$ of the optical cavity is high enough,
the non-linear interaction term is ``linearized'';
we may introduce shifted and rotated versions
$a, b$ of the annihilation and creation operators,
describing oscillations around the steady state:
\be
\label{eq:om:opshift}
\begin{aligned}
  \ha &= e^{i(\eta-\phi_L)} (a +s\I),\\
  \hb &= e^{i\zeta} (b +r\I),
\end{aligned}
\ee
where $\eta,\zeta$ are rotation angles and $s,r$ are complex shifts in the harmonic oscillator phase space,
all of them unspecified for now.
Moreover, we introduce the hatless position and momentum operators $(Q,P)$ and $(q,p)$ based on~$a,b$.
This yields
\be
\begin{aligned}
\ha^\dagger \ha &=
a^\dagger a +\re(s)Q +\im(s)P  +|s|^2\I,\\
\hb+\hb^\dagger &=
\cos(\zeta)q -\sin(\zeta)p +2\re(e^{i\zeta} r)\I,\\
\ha e^{i\phi_L} +\hc &=
\cos(\eta)Q -\sin(\eta)P +2\re(e^{i\eta} s)\I.
\end{aligned}
\ee
The cavity Lindblad operator $\sqrt{\kappa} \ha$ is equivalent to
$\sqrt{\kappa} a$ combined with the extra Hamiltonian term
\be
H_\text{Lind, cavity}/\hbar
= \frac{\kappa}{2} i(s^* a -s a^\dagger)
= \frac{\kappa}{2} (-\re(s)P +\im(s)Q),
\ee
and the oscillator Lindblad operators to
$\{\sqrt{\gamma'}b, \sqrt{\gamma' x}b^\dagger\}$ plus
the extra Hamiltonian term
\be
H_\text{Lind, osc}/\hbar
= (1-x) \frac{\gamma'}{2} i(r^* b -r b^\dagger)
= (1-x) \frac{\gamma'}{2} (-\re(r)p +\im(r)q).
\ee
Now the optomechanical Hamiltonian, expressed in terms of the transformed operators and including the Lindblad-induced terms above, is
\be
\begin{aligned}
\Hco''/\hbar
=\:&
(\Hco' +H_\text{Lind, cavity} +H_\text{Lind, osc})/\hbar\\
=\:&
(-\Delta -\gco(\cos(\zeta)q -\sin(\zeta)p +2\re(e^{i\zeta} r))) (a^\dagger a +\re(s)Q +\im(s)P  +|s|^2\I)\\
&+\Omega_m (b^\dagger b +\re(r)q +\im(r)p)
+\frac{E}{2} (\cos(\eta)Q -\sin(\eta)P)\\
&+\frac{\kappa}{2} (-\re(s)P +\im(s)Q)
+(1-x) \frac{\gamma'}{2} (-\re(r)p +\im(r)q),
\end{aligned}
\ee
where we immediately dropped any terms that are mere multiples of identity.
Next, the unwanted interaction cross-terms $Pq$, $Qp$ and~$Pp$ are eliminated by choosing
$\sin(\zeta)=\im(s)=0$.
Thus $s$ is real, and $\zeta = 0$ since $\zeta = \pi$ would be just an uninteresting $q,p$ inversion.
We obtain
\be
\begin{aligned}
\Hco''/\hbar
=\:&
-\Delta' a^\dagger a
+\Omega_m b^\dagger b
-\gco s Qq
-\gco a^\dagger a q\\
&+Q(-\Delta' s +\frac{E}{2} \cos(\eta))
+P(-\frac{E}{2} \sin(\eta) -\frac{\kappa}{2}s)\\
&+q(-\gco |s|^2 +\Omega_m \re(r) +(1-x)\frac{\gamma'}{2} \im(r))\\
&+p(\Omega_m \im(r) -(1-x)\frac{\gamma'}{2} \re(r))
\end{aligned}
\ee
where the shifted detuning
$\Delta' := \Delta +2\gco\re(r) = \wl-\wc'$,
and the shifted cavity frequency
$\wc' := \wc -2\gco\re(r)$.
We can see that the shift~$s$ acts as an \emph{enhancement factor} on the
linear cavity-oscillator interaction term~$-\gco s Qq$.
The remaining linear terms can be eliminated by fixing the remaining free parameters
$\eta, s, r$ such that
\be
\begin{aligned}
\frac{E}{2} \cos(\eta) &= (\Delta+2\gco\re(r))s\\
\frac{E}{2} \sin(\eta) &= -\frac{\kappa}{2}s\\
\gco s^2 &= \Omega_m \re(r) +(1-x)\frac{\gamma'}{2} \im(r)\\
\Omega_m \im(r) &= (1-x)\frac{\gamma'}{2} \re(r)
\end{aligned}
\ee
or
\be
\begin{aligned}
\gco s^2 &= \left(\Omega_m +\frac{(1-x)^2}{\Omega_m}\left(\frac{\gamma'}{2}\right)^2\right) \re(r)\\
\left(\frac{E}{2}\right)^2  &= \left((\Delta+2\gco\re(r))^2 +\left(\frac{\kappa}{2}\right)^2\right)s^2
\end{aligned}
\ee
This yields a cubic equation for $\re(r)$.
If we approximate $\gamma' \ll \Omega_m$ (given in nearly all cavity optomechanics realizations), the oscillator shift is
$r = \frac{\gco s^2}{\Omega_m}$.
If we instead assume the coupling enhancement factor $s > 0$ given and treat the driving laser amplitude~$E$ as a free parameter,
we may easily
solve $r$ and $\Delta'$, and then~$E$ and~$\eta$.
This way we obtain $s = |\alpha'|$ and $\eta = \arg(\alpha')$ for the transformed coherent state parameter (cf. Eq.~\eqref{eq:alpha})
\be
\alpha' = \frac{E/2}{\Delta' +i\kappa/2}.
\ee
We thus end up with the relatively simple Hamiltonian
(plus the counter-rotating term)
\begin{align}
\label{eq:om:co_rotframe}
\Hco''/\hbar
=&
-\Delta' a^\dagger a
+\Omega_m b^\dagger b
-\gco s \: Qq
-\gco a^\dagger a q
+\frac{E}{2}(a e^{-i(2\wl t +2\phi_L-\eta)} +\hc).
\end{align}

Next, we perform the same operator substitutions to the atomic Hamiltonian,
again expressing $\ha$ and~$\hb$ in terms of the hatless versions using Eqs.~\eqref{eq:om:opshift},
together with the further substitutions
\be
\begin{aligned}
\phi &= \phi_a +\phi_c +\eta -\phi_L,\\
\sigma_+ &= e^{i\phi}\hs,\\
\phi_R'(t) &= \phi_R(t) +\phi,
\end{aligned}
\ee
yielding
\begin{align}
\label{eq:om:atom_rotframe}
\Ha''/\hbar
=\:&
(\wa-\wr) \sigma_+ \sigma_-
+\frac{\ac}{2}(\sigma_+ (e^{-i\phi_R'}+e^{i(2\wr t +\phi_R'-2\phi)}) +\hc)\\
\notag
&+\gac ((a +s\I) \sigma_+ e^{i(\wr-\wl) t} +(a^\dagger+s\I) \sigma_+ e^{i((\wl+\wr) t -2\phi+2\phi_a)} +\hc)\\
\notag
=\:&
(\wa-\wr) \sigma_+ \sigma_-
+\gac (a\sigma_+ e^{i(\wr-\wl) t} +a^\dagger \sigma_+ e^{i((\wl+\wr) t -2\phi+2\phi_a)} +\hc)\\
\notag
&\:+\left[\left(\frac{\ac}{2}(e^{-i\phi_R'}+e^{i(2\wr t +\phi_R'-2\phi)}) +\gac s (e^{i(\wr-\wl) t} +e^{i((\wl+\wr) t -2\phi+2\phi_a)})\right)\sigma_+ +\hc\right].
\end{align}
Since the new hatless atomic raising and lowering operators are simply phase-rotated versions
of the originals, no extra Hamiltonian terms are induced by the Lindblad dissipator.

The phases~$\phi_L$, $\phi_a$ and~$\phi_c$
were absorbed into the transformed operators and the control phase~$\phi_R'$,
and only remain in the counter-rotating terms (which we approximate away as they perturb the dynamics only slightly).
The non-linear term $-\gco a^\dagger a q$ in $\Hco''$ is also typically very weak and can be ignored
in our weak coupling scenario, i.e., $\gco\ll(\kappa,\Omega_m)$.
In Table~\ref{table:significance} we show significance estimates for all the discarded terms.

The dynamics (in the rotating frame) given by $\Hco''+\Ha''$ together with the Lindblad dissipator
$D{\{\sqrt{\kappa}a, \sqrt{\gamma'}b, \sqrt{\gamma' x}b^\dagger\}}$
are equivalent to $\Hco'+\Ha'$ together with
$D{\{\sqrt{\kappa}\ha, \sqrt{\gamma'}\hb, \sqrt{\gamma' x}\hb^\dagger\}}$,
but expressed in terms of the new,
rotated and shifted operators $a$ and~$b$, which fulfill the original bosonic commutation relations.
In terms of the eigenstates of the transformed number operator~$a^\dagger a$, if $\gco=0$ the cavity steady state is~$\ket{0}$,
and with realistic enhanced coupling strengths it remains close to~$\ket{0}$.
We have
\be
\expect{\ha} = \bra{0}\ha\ket{0} = e^{i(\eta-\phi_L)} \bra{0}(a+s\I)\ket{0}
= e^{i(\eta-\phi_L)} s
= e^{-i\phi_L} \alpha'.
\ee
The operator shift~\eqref{eq:om:opshift}
thus enables us to truncate the computational
Hilbert space much more heavily, even when $|s|$ is large.
From now on we always use the shifted-and-rotated operators and their eigenstates.

\subsection{Steady state}
\label{sec:steady_state}

The full system Hamiltonian, after dropping the counter-rotating terms
in Eqs.~\eqref{eq:om:co_rotframe} and~\eqref{eq:om:atom_rotframe}, is
\begin{align}
\label{eq:om:full}
\notag
H''/\hbar
=& (\Hco'' +\Ha'')/\hbar
=
(\wc'-\wl) a^\dagger a
+\Omega_m b^\dagger b
+(\wa-\wr) \sigma_+ \sigma_-
-\gco s \: Qq
-\gco a^\dagger a q\\
&+\gac (a\sigma_+ e^{i(\wr-\wl) t} +\hc)
+\left[\left(\frac{\ac}{2} e^{-i\phi_R'} +\gac s e^{i(\wr-\wl) t}\right)\sigma_+ +\hc\right].
\end{align}
Depending on whether we want a two-mode squeezing or a hopping interaction,
we choose the laser-cavity detuning
$\Delta' = \wl-\wc' = \pm \Omega_m$.

In the absence of atomic control, $\ac = 0$, $\wr$ is an arbitrary constant, and we may choose $\wr = \wl$ to obtain
\begin{align}
\notag
H''/\hbar
=&
-\Delta' a^\dagger a
+\Omega_m b^\dagger b
+(\wa-\wl) \sigma_+ \sigma_-
-\gco s \: Qq
-\gco a^\dagger a q\\
&+\gac (a\sigma_+ +\hc)
+\gac s \left(\sigma_+ +\hc\right)
\end{align}
The presence of the atom modifies the steady state
into which the system evolves during an initial period of laser driving of the cavity.
The strong $\gac s (\sigma_+ +\hc)$ term makes the steady state impure, unless
the atom is far detuned from the cavity in which case the system ends up
close to the ground state (of the transformed operators),
as the oscillator is cooled by the hopping interaction with the cavity.
With this assumption, with \lecocqII{} parameters we obtain a steady state
with the cavity populations $p_0 = 0.9922, p_1 = 0.0078$ and the oscillator populations
$p_0 = 0.9912, p_1 = 0.0087$.

At the start of the control sequence, $t=0$, we transform the steady state to the simulation frame.
Since the frames coincide at this point, this does nothing to the state.

\subsection{Control system}\label{app:controlsystem}

If $\wr \neq \wl$,
in order to obtain a constant drift Hamiltonian,
we need one more rotating frame transformation to stop the rotation of the atom-cavity interaction term
while keeping either the two-mode squeezing or the hopping interaction term fixed.
This is accomplished using the generator
$H_0/\hbar = (\wr-\wl) a^\dagger a \mp(\wr-\wl) b^\dagger b$
(in terms of the transformed operators),
which yields
\begin{align}
\label{eq:om:sim}
\notag
H'''/\hbar
=&
\delta_R' (-a^\dagger a \pm b^\dagger b)
+(\wa-\wr) \sigma_+ \sigma_-
-\gco s \: (abe^{-i(\wr-\wl)(1\mp 1)t} +ab^\dagger e^{-i(\wr-\wl)(1\pm 1)t} +\hc)\\
&+\gac (a\sigma_+ +\hc)
+\left[\left(\frac{\ac}{2} e^{-i\phi_R'} +\gac s e^{i(\wr-\wl)t}\right)\sigma_+ +\hc\right]
\end{align}
where $\delta_R' = \wr-\wc'$.
\newpage
The $\gac s$ term in the above equation, resulting from the shifted part of the atom-cavity interaction term,
is somewhat problematic. We propose three possible strategies for dealing with it:
\begin{itemize}
\item
Actively cancel it using the control signal $(\ac, \phi_R'(t))$ produced by the signal generator.
For this strategy we need a high Rabi frequency for the control signal, and a high sample rate
for the signal generator.
\item
Passively cancel it using another harmonic signal on top of $\ac$,
which also requires a high Rabi frequency for the canceling signal.
Such a strong driving has been, e.g., used in Ref.~\cite{baur_measurement_2009}.
\item
Include it in the simulation and optimization.
To have a fixed $H_\text{drift}$ we need to set $\wr=\wl$.
Since $\wl$ is not that far from cavity resonance,
this may weaken the control system.
\end{itemize}

We choose the hopping interaction by driving the cavity with a red-detuned laser, with the laser-cavity detuning
$\Delta' = \wl-\wc' = -\Omega_m$.
Dropping the counter-rotating $ab$ interaction term,
Eq.~\eqref{eq:om:sim} yields the time-independent drift Hamiltonian
\begin{align}
H_\text{drift}/\hbar
=&
-\delta_R' (a^\dagger a +b^\dagger b)
+({\wa}_0-\wr) \sigma_+ \sigma_-
-\gco s \: (ab^\dagger +\hc) +\gac (a\sigma_+ +\hc).
\end{align}
In our control scheme, the atom resonance frequency
$\wa = {\wa}_0 +2\pi u_{\text{detuning}}(t)$
is split into a constant part and a tunable part.

The remaining terms in Eq.~\eqref{eq:om:sim} constitute the time-dependent control Hamiltonian.
The atomic control signal is split into $X$ and $Y$ components
$u_{\text{atomX}}(t) = \frac{\ac}{2\pi} \cos(\phi_R'(t))$ and
$u_{\text{atomY}}(t) = \frac{\ac}{2\pi} \sin(\phi_R'(t))$
that can be independently adjusted:
\begin{align}
\notag
H_\text{control}(t)/\hbar
&=
u_{\text{detuning}}(t) \: 2\pi \sigma_+ \sigma_-
+\frac{\ac}{2} \left[e^{-i\phi_R'}\sigma_+ +e^{i\phi_R'}\sigma_-\right]\\
\notag
&=
u_{\text{detuning}}(t) \: 2\pi \sigma_+ \sigma_-
+\frac{\ac}{2} \left[\cos(\phi_R')(\sigma_++\sigma_-) +\sin(\phi_R')(-i)(\sigma_+ -\sigma_-)\right]\\
&=
u_{\text{detuning}}(t) \: 2\pi \sigma_+ \sigma_-
+u_{\text{atomX}}(t) \: \pi (\sigma_++\sigma_-)
+u_{\text{atomY}}(t) \: \pi (-i)(\sigma_+-\sigma_-).
\end{align}
The $2\pi$ factors were introduced to make the control fields $u(t)$ normal frequencies.
The dissipation processes are described using the Lindblad operators
$\{\sqrt{\kappa}a, \sqrt{\gamma'}b, \sqrt{\gamma' x}b^\dagger, \sqrt{\kappa_a} \sigma_-\}$.

\newpage
\phantom{X}
\bigskip\bigskip

\begin{table*}[h]
\renewcommand*{\arraystretch}{0.8}
\begin{tabular}{llrccl}
  \hline \hline
  symbol & meaning & & \lecocqII & \lecocqIII\\
  \hline \hline
  $\sigma_-$ & transformed atom annihilation operator\\
  $a$ & transformed cavity annihilation operator\\
  $b$ & transformed oscillator annihilation operator\\
  $s$   & cavity shift (boosts the linearized $\gco$ coupling) && 100 & 120\\
  $\re(r)$ & oscillator shift, real part && $7.5$ & $2.7$\\
  $\im(r)$ & oscillator shift, imaginary part && $3.6$ & $1.3$ & $\cdot 10^{-5}$\\
  $\wa$ & atom resonance frequency & $2\pi \cdot$ & \multicolumn{2}{c}{9--13.5} & GHz\\
  $\wc$ & cavity resonance frequency & $2\pi \cdot$ & \multicolumn{2}{c}{10.188} & GHz\\
  $\Omega_m$ & oscillator resonance frequency & $2\pi \cdot$ & \multicolumn{2}{c}{15.9} & MHz\\
  $\gac$ & atom-cavity coupling & $2\pi \cdot$ & \multicolumn{2}{c}{12.5} & MHz\\
  $\gco$ & cavity-oscillator coupling & $2\pi \cdot$ & $12$ & $3$ & kHz\\
  $\gac s$ & boosted atom-cavity coupling & $2\pi \cdot$ & $1.25$ & $1.50$ & GHz\\
  $\gco s$ & boosted cavity-oscillator coupling & $2\pi \cdot$ & $1.2$ & $0.36$ & MHz\\
  $\kappa_a$ & atom decay rate & $2\pi \cdot$ & \multicolumn{2}{c}{1} & MHz\\
  $\kappa$   & cavity decay rate & $2\pi \cdot$ & $1$ & $0.2$ & MHz\\
  $\gamma$   & oscillator decay rate & $2\pi \cdot$ &  \multicolumn{2}{c}{150} & Hz\\
  $\wl$ & cavity-driving laser frequency\\
  $E$ & cavity-driving laser amplitude & $2\pi \cdot$ & 3.18 & 3.82 & GHz\\
  $\phi_L(t)$ & cavity-driving laser phase\\
  $\wr$ & atom control frequency\\
  $\ac \lessapprox \frac{E}{100}$ & atom control amplitude & $2\pi \cdot$ & 32 & 38 & MHz\\
  $\phi_R'(t)$ & atom control phase\\
  $-2\gco\re(r)$ & cavity resonance frequency shift & $2\pi \cdot$ & $-0.18$ & $-0.016$ & MHz\\
  $\wc' = \wc-2\gco\re(r)$ & shifted cavity resonance frequency & $2\pi \cdot$ & \multicolumn{2}{c}{10.188} & GHz\\
  $\Delta = \wl-\wc$      & laser detuning\\
  $\Delta' = \wl-\wc'$    & shifted laser detuning &&  \multicolumn{2}{c}{$-\Omega_m$}\\
  $\delta_a' = \wa-\wc'$  & shifted atom detuning\\
  $\delta_R' = \wr-\wc'$  & shifted atomic control detuning\\
  $T$ & temperature && $25$ & $10$ & mK\\
  $e^{-\frac{\hbar\wa}{kT}}$ & atom Boltzmann factor & & $\sim 3 \cdot 10^{-8}$ & $\sim 6 \cdot 10^{-21}$\\
  $e^{-\frac{\hbar\wc}{kT}}$ & cavity Boltzmann factor && $3.2 \cdot 10^{-9}$ & $5.8 \cdot 10^{-22}$\\
  $x = e^{-\frac{\hbar\Omega_m}{kT}}$ & oscillator Boltzmann factor && $0.97$ & $0.93$\\
  $\nn = x/(1-x)$ & expected number of oscillator phonons && $32.3$ & $12.6$\\
  $\gamma' = \gamma (\nn+1)$ & effective oscillator decay rate & $2\pi \cdot$ & $5.0$ & $2.0$ & kHz\\
  \hline \hline
\end{tabular}
\caption{
\label{table:symbols}
Summary of used symbols and system parameters.}
\end{table*}

\newpage
\phantom{X}
\bigskip\bigskip

\begin{table*}[h]
\setlength{\tabcolsep}{0.8em}
\begin{tabular}{lccc}
  \hline \hline
  measure & definition & \lecocqII & \lecocqIII\\
  \hline \hline
  sideband resolution & $\frac{\Omega_m}{\kappa}$ & 15.9 & 79.5\\
  cavity-oscillator cooperativity & $\frac{|\gco s|^2}{\kappa \, \gamma \, \nn}$ & 298 & 343\\
  cavity-oscillator coupling-dissipation ratio & $\frac{|\gco s|}{\max(\kappa, \gamma)}$ & 1.2 & 1.8\\
  atom-cavity cooperativity & $\frac{\gac^2}{\kappa \, \kappa_a}$ & 156 & 781\\
  atom-cavity coupling-dissipation ratio & $\frac{\gac}{\max(\kappa, \kappa_a)}$ & 12.5 & 12.5\\
  \hline \hline
\end{tabular}
\caption{
\label{table:measures}
Important parameter ratios for the hybrid optomechanical system.
Both parameter sets place us in the high-cooperativity, strong-coupling, resolved-sideband regime.
}
\end{table*}

\bigskip\bigskip\bigskip\bigskip

\newcommand{\xxx}{\min(\wa)}

\begin{table*}[h]
\setlength{\tabcolsep}{0.8em}
\begin{tabular}{lccc}
\hline \hline
term & significance & \lecocqII & \lecocqIII\\
\hline \hline
driving laser, co-rotating       & $\frac{E}{2|\wl-\wc|}$  
& 99 & 120\\
\hline
driving laser, counter-rotating  & $\frac{E}{2|\wl+\wc|}$ & 0.078 & 0.094\\
atomic control, counter-rotating & $\frac{R}{2|\wr+\xxx|}$ & 0.00082 & 0.00098\\
$\gco$, non-linear part     & $\frac{\gco}{\Omega_m}$ & 0.00075 & 0.00019\\
$\gco s$, two-mode squeezing & $\frac{|\gco s|}{|2\Delta'|}$ & 0.038 & 0.011\\
$\gac$, counter-rotating    & $\frac{\gac}{|\xxx+\wc'|}$ & 0.00063 & 0.00063\\
$\gac s$, counter-rotating & $\frac{|\gac s|}{|\xxx+\wl|}$ & 0.063 & 0.076\\
\hline \hline
\end{tabular}
\caption{
\label{table:significance}
Significance estimates for various Hamiltonian terms.
The driving laser co-rotating term is shown for comparison.
All the terms in the lower part of the table are discarded in rotating wave approximations.
}
\end{table*}

\end{document}